\documentclass[final,times,twocolumn]{elsarticle}
\usepackage{graphicx,amssymb,amsmath,hyperref}
\usepackage{multirow,dcolumn,bm,latexsym,soul}

\journal{Physics of the Dark Universe}
\begin{document}
\newcommand{\be}{\begin{equation}}
\newcommand{\ee}{\end{equation}}
\newcommand{\bq}{\begin{eqnarray}}
\newcommand{\eq}{\end{eqnarray}}
\begin{frontmatter}

\title{Impact of spatial curvature on forecast constraints from standard and differential redshift drift measurements}
\author[inst1,inst2]{C. J. A. P. Martins\corref{cor1}}\ead{Carlos.Martins@astro.up.pt}
\author[inst1,inst3]{M. A. F. Melo e Sousa}\ead{up202004740@edu.fc.up.pt}
\author[inst4]{S. Q. Fernandes}\ead{sofiafernandes2910@gmail.com}
\author[inst1,inst2,inst3]{C. M. J. Marques}\ead{Catarina.Marques@astro.up.pt}
\address[inst1]{Centro de Astrof\'{\i}sica da Universidade do Porto, Rua das Estrelas, 4150-762 Porto, Portugal}
\address[inst2]{Instituto de Astrof\'{\i}sica e Ci\^encias do Espa\c co, CAUP, Rua das Estrelas, 4150-762 Porto, Portugal}
\address[inst3]{Faculdade de Ci\^encias, Universidade do Porto, Rua do Campo Alegre, 4150-007 Porto, Portugal}
\address[inst4]{Escola B\'asica e Secund\'aria de Ermesinde, Praceta D. Ant\'onio Ferreira Gomes, 4445-398 Ermeside, Portugal}
\cortext[cor1]{Corresponding author}

\begin{abstract}
{The redshift drift of objects following the cosmological expansion is a unique model-independent probe of background cosmology, detectable by astrophysical facilities presently under construction. Previous forecasts for such measurements assume flat universes. We explore the impact of relaxing this assumption on the constraining power of the redshift drift, focusing on the two most promising routes for its measurement: the SKA at low redshifts, and the Golden Sample for the ELT's ANDES spectrograph at higher redshifts. We also discuss the cosmological sensitivity of possible differential redshift drift measurements, both on their own and, for the specific case of the Golden Sample, in combination with the standard method. Overall, we find that the sensitivity of the redshift drift to curvature is comparable to that of matter (especially at low redshifts) and higher than the sensitivity to the dark energy equation of state. We also show that the sensitivity of redshift drift measurements to these cosmological parameters is asymmetric with respect to the curvature parameter, being different for open and closed universes with the same absolute value of the curvature parameter $\Omega_k$.}
\end{abstract}
\begin{keyword}
Observational cosmology \sep Redshift drift \sep Spatial curvature \sep Fisher Matrix forecasts
\end{keyword}
\end{frontmatter}

\section{Introduction}
\label{introd}

All astrophysical measurements, including cosmological ones, are, to some extent, model-dependent, in the sense of requiring an underlying model for interpretation. For example, a measurement of the cosmological matter density is the measurement of the parameter $\Omega_m$ of a model which is assumed to be the correct one, and typically assumes a homogeneous and isotropic universe, the validity of General Relativity, and possibly more. One may therefore ask what is the least model-dependent cosmological observation that one could conceivably do. The answer is the redshift drift of objects following the background cosmological expansion, also known as the Sandage test \cite{Sandage}, 

Conceptually, the idea is beautifully simple: the redshift of such an object changes in time, and if one can detect this change one is (to put it simply) seeing the universe expand in real time. This is fundamentally different from all traditional astrophysical observations: in those one passively maps our past light cone, while with the redshift drift one directly compares different past light cones. The implications of its detection are equally fundamental: just as detecting redshift is evidence for cosmological expansion, detecting its drift is evidence of nonuniform expansion---in other words, of deceleration or acceleration, depending on the sign of this drift. In particular, a direct measurement of a positive drift signal implies a violation of the strong energy condition, and hence the presence of some form of dark energy (be it a cosmological constant or some alternative mechanism) accelerating the Universe \citep{Liske,Uzan,Quercellini,Heinesen}.

The practical difficulty is that the characteristic timescale of the cosmological expansion is far larger than that of typical observations or experiments. The value of the expected spectroscopic velocity signal will depend on the redshift at which it is measured, but it is typically of the order of a few centimetres per second per decade, leading, among others, to extremely stringent requirements on the precision and stability of astrophysical facilities able to detect it.

The best presently available upper limits on this signal, obtained in the optical and radio bands, are three orders of magnitude larger than the expected signal \cite{Darling,Cooke}, and limited by systematics. An ongoing experiment (PI: C. Martins), with the ESPRESSO spectrograph and an experiment time of about one year, will significantly improve these limits, but a detection of the signal is only within the reach of the next generation of astrophysical facilities, which are presently under construction. Specifically, a detection is within the reach of the Square Kilometre Array (SKA) at low redshifts $z\lesssim 1$ \cite{Klockner} (in its full configuration, not reduced versions thereof) and of the ANDES spectrograph at the ELT at higher redshifts $2\lesssim z \lesssim 5$ \cite{Liske,HIRES,ANDES}. We note that by probing different redshifts the two facilities optimally complement each other. On the one hand, redshift drift measurements at high redshift can have larger broader impacts, since high-redshift observations are comparatively more scarce. On the other hand, as has already been mentioned, a direct measurement of a positive drift signal, which for observationally plausible models is possible only at low redshifts, demonstrates the presence of some form of dark energy.

Although in principle the redshift drift measurements are model-independent, in practice they will of course be used to place constraints on relevant plausible models and their parameters, and, in particular, they will be combined with the traditional cosmological datasets. The usual forecasting tools can be used to assess the cosmological impact of such measurements, which is also helpful in optimizing future observing strategies. Several detailed forecasts, both based on Fisher Matrix and MCMC techniques, have recently been presented \cite{Alves,Rocha,SKA}. However, one aspect that so far has been neglected is the possible impact of curvature. One exception is \cite{Balbi}, where the effect of curvature is qualitatively illustrated in plots, but no quantitative forecasts are provided.

While most astrophysical observations, including Planck \cite{Planck}, constrain curvature to be small, this result is sometimes disputed, e.g. in \cite{Closed} whose analysis suggests a closed universe. This  motivates an extension of previous analyses allowing for non-vanishing curvature. Here we report the results of this analysis. We start with the standard redshift drift, for which plausible scenarios for SKA and ELT observations are broadly understood, including possible observational strategies, which take into account practical constraints \cite{Cristiani,Dong}. We then extend our analysis to the more recently proposed differential redshift drift \cite{Cooke}, first with a generic (i.e., conceptual) approach, but then focusing on the ELT, for which such a measurement is a realistic possibility (unlike for the SKA). In particular, we provide the first forecast constraints on the joint measurements of the standard and differential redshift drift, which is a realistic possibility for the quasars in the ANDES Golden Sample, and also comment on the advantages of such joint measurements.

\section{Standard redshift drift}
\label{stddri}

The redshift drift of an astrophysical object following the cosmological expansion, for someone observing it over a time span $\Delta t$, is \cite{Sandage}
\be
\frac{\Delta z}{\Delta t}=H_0 \left[1+z-E(z)\right]\,,
\ee
although the measurement will effectively be done in velocity space, in which we have
\be\label{specvel}
\Delta v=\frac{c\Delta z}{1+z}=(cH_0\Delta t)\left[1-\frac{E(z)}{1+z}\right]\,.
\ee
Here we have defined the dimensionless Hubble parameter, $E(z)=H(z)/H_0$; it is also convenient to define a dimensionless redshift drift
\be
S_z=\frac{1}{H_{100}}\frac{\Delta z}{\Delta t}=h\left[1+z-E(z)\right]\,,
\ee
where $H_0=hH_{100}$ and $H_{100}=100$ km/s/Mpc. The corresponding spectroscopic velocity is
\be
S_v={\Delta v}=k_dh\left[1-\frac{E(z)}{1+z}\right]\,,
\ee
where we further defined $k_d=cH_{100}\Delta t$ (not to be confused with the curvature parameter). This is a constant parameter, for a given observation time, with units of cm/s. Specifically, for an experiment time $\Delta t=1$ year we have\footnote{Under the assumption that one year consists of 365.25 days; if one assumes 365, then $k_d=3.064$ cm/s.} $k_d=3.066$ cm/s.

\begin{figure*}
\begin{center}
\includegraphics[width=0.49\textwidth]{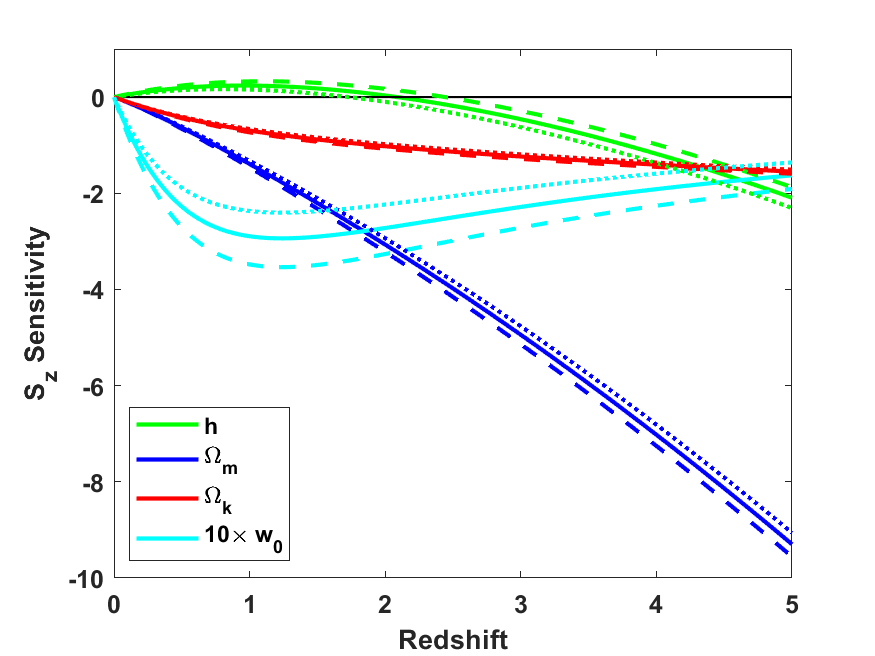}
\includegraphics[width=0.49\textwidth]{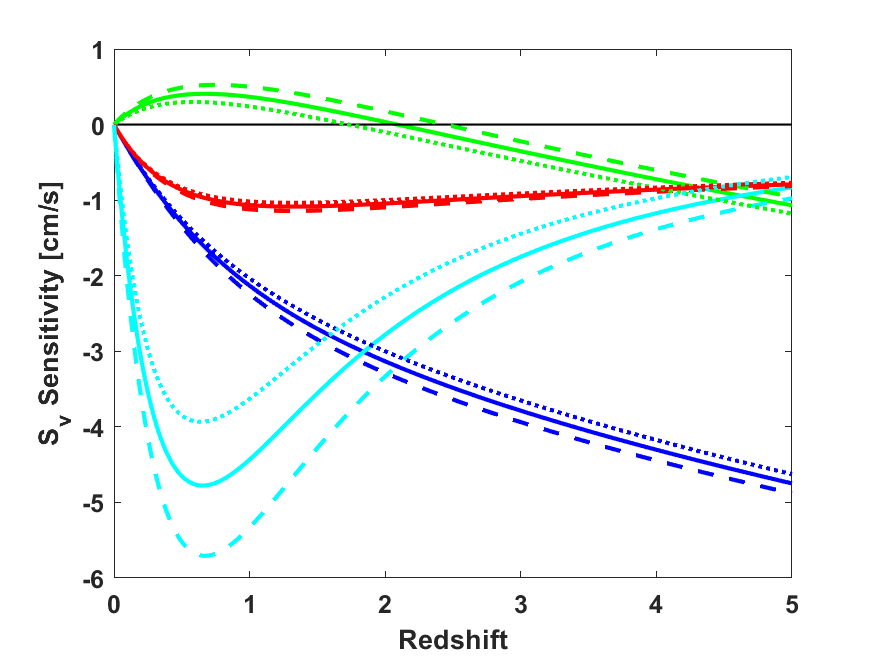}
\end{center}
\caption{Theoretical sensitivities of the redshift drift (left panel, in dimensionless units) and the spectroscopic velocity (right panel, in units of cm/s) to the cosmological parameters in the $w_0$CDM parameterization, for three different fiducial models. Note that for plotting convenience the sensitivity of $w_0$ has been multiplied by a factor of ten. The zero sensitivity line is also shown in black. The solid, dashed, and dotted lines correspond to $\Omega_k=0$, $\Omega_k=-0.1$, and $\Omega_k=+0.1$ respectively.}
\label{figure01}
\end{figure*}

In our forecasts we assume a class of homogeneous and isotropic universes comprising matter, dark energy with a constant equation of state, and possibly curvature,
\be\label{fried}
E^2(z)=\Omega_k(1+z)^2+\Omega_m(1+z)^3+(1-\Omega_m-\Omega_k)(1+z)^{3(1+w_0)}\,;
\ee
we ignore the radiation component since we are dealing with low-redshift observations throughout.

Figure \ref{figure01} shows, for the four cosmological parameters $p_i$, the theoretical sensitivity coefficients\footnote{ We emphasize that these are theoretical sensitivity coefficients, to the parameters of the fiducial model. They have no relation to particular instrumental configurations.} of the redshift drift ($\partial S_z/\partial p_i$) and the spectroscopic velocity ($\partial S_v/\partial p_i$), for three fiducial models with $\Omega_k=0,\pm0.1$, and having in common the three other parameters: $h=0.7$, $\Omega_m=0.3$, and $w_0=-1$. Note that the two sensitivity coefficients are related via
\be\label{senscoef}
\frac{\partial S_v}{\partial p_i}=\frac{k_d}{1+z} \frac{\partial S_z}{\partial p_i}\,.
\ee

A first interesting observation is that at low redshifts (below $z\approx 2$) there is a higher sensitivity to closed universes than to open ones for all four parameters, while for higher redshifts the behaviour will depend on how one defines an overall sensitivity. A second one is that in terms of $S_z$ the sensitivity to $\Omega_k$ is akin to that of the matter density, in the sense that they are both monotonically decreasing functions (with the sensitivity to matter being larger in absolute value). On the other hand, in terms of $S_v$ the sensitivity to $\Omega_k$ is more similar to that of the dark energy equation of state, as both have a maximum (again, in absolute value) at $z\approx 1$, with the sensitivity to curvature being the stronger one.

\begin{figure*}
\begin{center}
\includegraphics[width=0.49\textwidth]{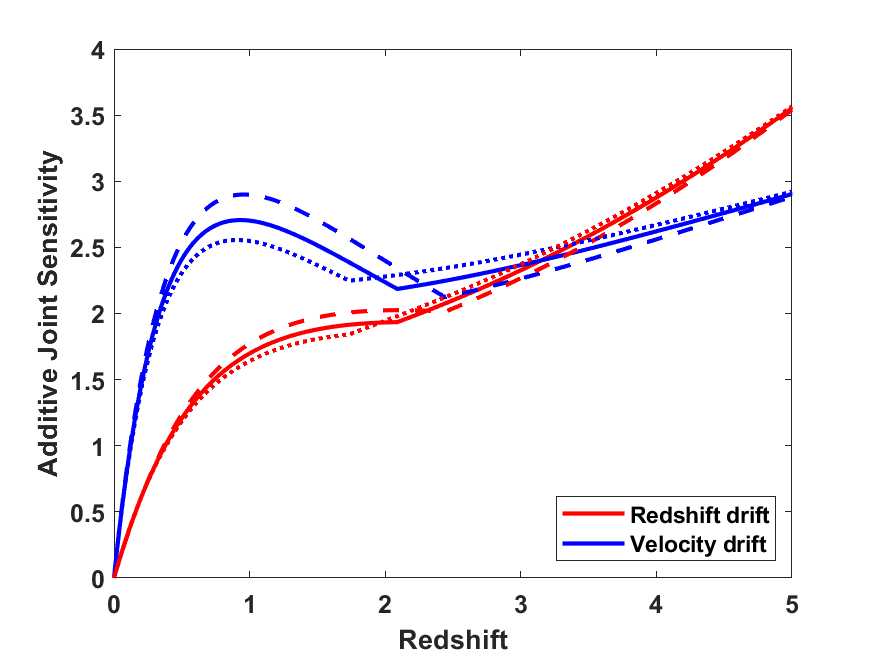}
\includegraphics[width=0.49\textwidth]{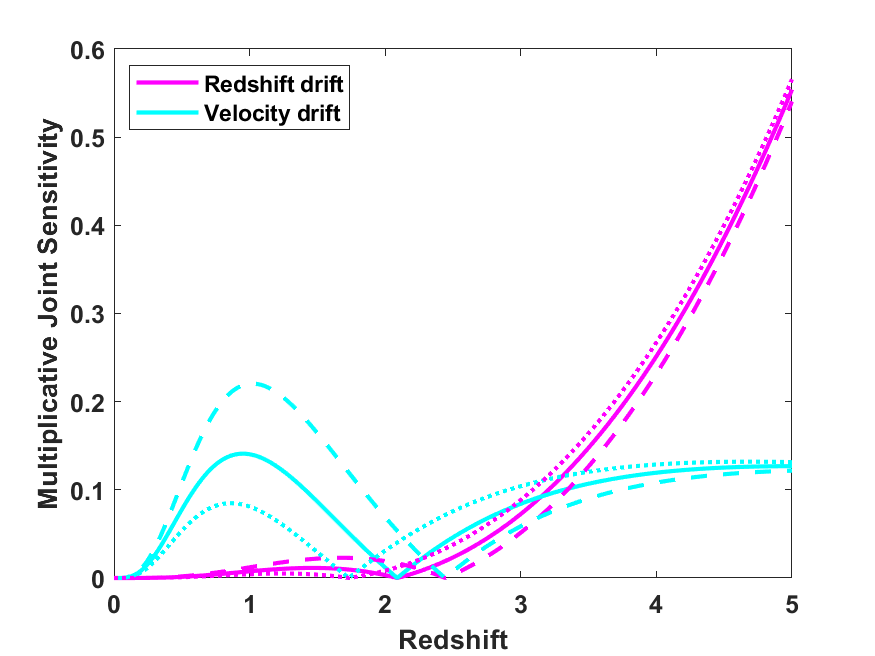}
\end{center}
\caption{Overall additive (left panel) and multiplicative (right panel) sensitivities, for the redshift drift (red/magenta curves) and velocity (blue/cyan curves). The solid, dashed, and dotted lines correspond to $\Omega_k=0$, $\Omega_k=-0.1$, and $\Omega_k=+0.1$ respectively.}
\label{figure02}
\end{figure*}

A purely phenomenological attempt at defining overall sensitivities is depicted in Figure \ref{figure02}. One can take the absolute values of each of the four sensitivities, normalized to their maximum values in the redshift range under consideration ($0\le z\le5$), and consider, in each case, either the sum or the product of these four sensitivities. Note that this entails the choice of giving equal weight to determinations of the fiducial model's four free parameters, which is certainly debatable. With this caveat, these joint sensitivities show that at low redshift the sensitivity is largest to closed universes, while at higher redshifts it is largest for open universes, but the impact of curvature in this case is much smaller than at lower redshifts. Moreover, if one wants to maximize this overall sensitivity, doing the measurements at the highest redshifts is always preferred, although if the goal is to maximize the spectroscopic velocity signal doing it at $z\approx 1$ is a comparable alternative. Indeed, $z\approx 1$ is the redshift for which, according to this somewhat arbitrary metric, the sensitivity to curvature is maximal.

Beyond the choice of a specific overall sensitivity, the important conclusion to be drawn from this analysis is that there are qualitatively different sensitivities at low and high redshifts, simply due to the fact that for our assumed $\Lambda$CDM fiducial model the drift signal is positive and negative, respectively, at those redshifts. Different choices of an overall sensitivity will lead to different quantitative numerical values, but do not change this qualitative overall behavior.

In practice, the conceptual analysis in the previous paragraph will have to be convolved with observational parameters such as the instrument sensitivity, signal-to-noise, redshift of the target, as well as scheduling constraints. Broadly speaking, for the SKA the expectation \cite{Klockner} is that the spectroscopic velocity uncertainty will increase with redshift (due to the decreasing number of sources), while for the ELT this trade-off is more complex \cite{Liske}: higher redshifts generally imply fainter sources, but they also enable more of the Lyman-$\alpha$ forest (upon which the measurement relies) to be observed. These aspects are taken into account in the specific forecasts we discuss in what follows.

\section{Standard redshift drift forecasts}
\label{stdfor}

We now use an extended version of the publicly available Fisher Matrix based FRIDDA\footnote{Available at \url{https://github.com/CatarinaMMarques/FisherCosmology}.} code \cite{FRIDDA}, further validated by a second independent code (written in Matlab instead of Python), to quantify the impact of adding curvature as an additional parameter on the forecasts of the cosmological impact of redshift drift measurements.

For a set of M model parameters $(p_1, p_2, ..., p_M)$ and model predictions for N observables $(f_1, f_2, ..., f_N)$, the Fisher matrix is defined as
\be
F_{ij}=\sum_{a=1}^N\frac{\partial f_a}{\partial p_i}\frac{1}{\sigma^2_a}\frac{\partial f_a}{\partial p_j}\,.
\ee
Previously known uncertainties on the parameters, known as priors, can be trivially added to the calculated Fisher matrix. In our case the observables are the various measurements of the spectroscopic velocity. As an illustration, consider the Fisher matrix for two model parameters, x and y; its inverse is the covariance matrix, 
\be
[F]^{-1}\equiv [C]=\begin{bmatrix}
\sigma^2_{x} \hfill & \sigma^2_{xy} \hfill \\
\sigma^2_{xy} \hfill & \sigma^2_{y}  \hfill
\end{bmatrix}\,,
\ee
where $\sigma^2_{x}$ and $\sigma^2_{y}$ are the uncertainties in the x and y parameters marginalizing over the other, while $\sigma^2_{xy}=\rho \sigma_{x}\sigma_{y}$ with $\rho$ being the correlation coefficient, ranging from $\rho=0$ for independent parameters to $\rho=\pm1$ for fully correlated and fully anticorrelated parameters. It's also useful to define a Figure of Merit
\be
FoM=\frac{1}{\sigma_x\sigma_y\sqrt{1-\rho^2}(\Delta\chi^2)}=\frac{1}{\left(\sigma^2_{x} \sigma^2_{y} - \sigma^4_{xy}  \right)^{1/2}(\Delta\chi^2)}\,,\label{FoMs}
\ee
which is proportional to the inverse of the area of the confidence ellipse of the two parameters. In the above, $\Delta\chi^2$ identifies the confidence interval of interest; for example, $\Delta\chi^2=2.3$ corresponds to the $68.3\%$ confidence level, which is the choice we use in what follows. This can straightforwardly be generalized to more than two parameters.

We will assume a set of redshift drift measurements from the SKA and the ELT, commensurate with current expectations for both facilities, and also used in other recent forecasts (which do not include the curvature parameter). For the former, following \cite{Klockner}, we assume a (possibly optimistic) set of ten measurements, uniformly spaced between redshifts $z = 0.1$ and $z = 1.0$ and with the associated spectroscopic velocity uncertainties equally spaced between $1\%$ and $10\%$ respectively. For the latter we assume the recently proposed Golden Sample \cite{Cristiani} of seven bright high-redshift quasars, spanning the redshift range $3.0\le z\le 4.8$, and with a total of 1500 hours of observations equally divided between the quasars, Note that this observation time is much smaller than the 4000 hours initially assumed in \cite{Liske} and in the forecasts therein.

\begin{figure*}
\begin{center}
\includegraphics[width=0.32\textwidth]{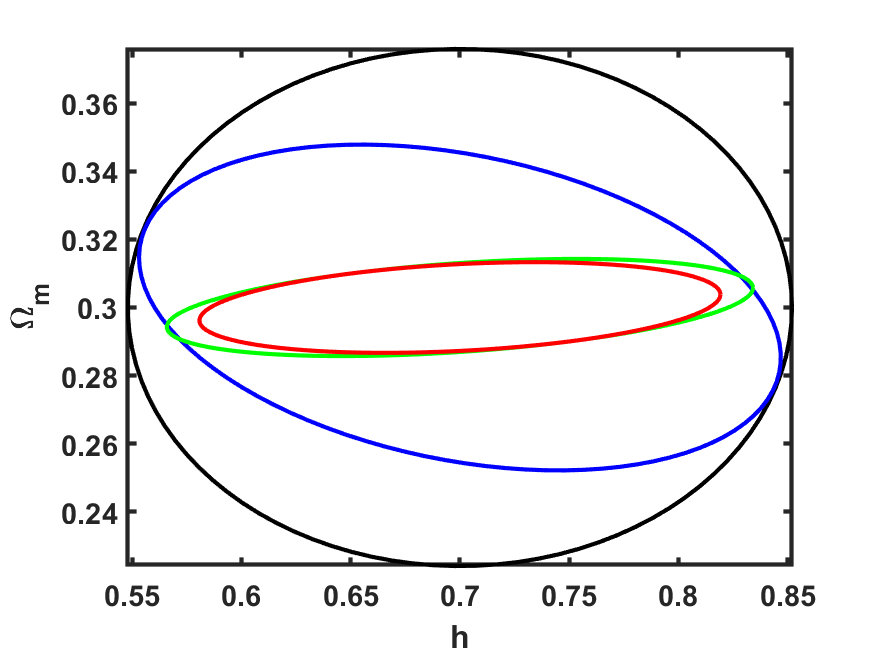}
\includegraphics[width=0.32\textwidth]{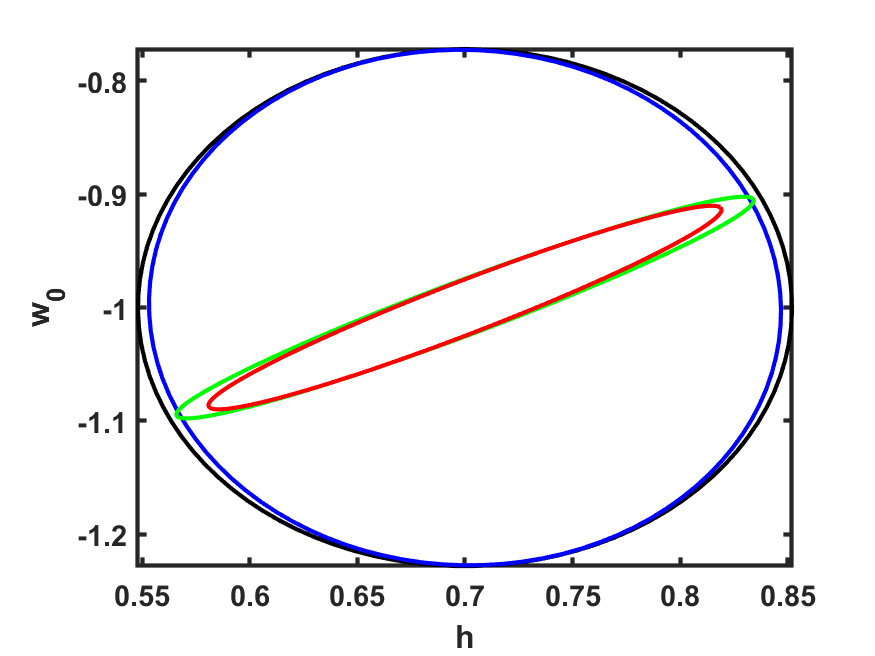}
\includegraphics[width=0.32\textwidth]{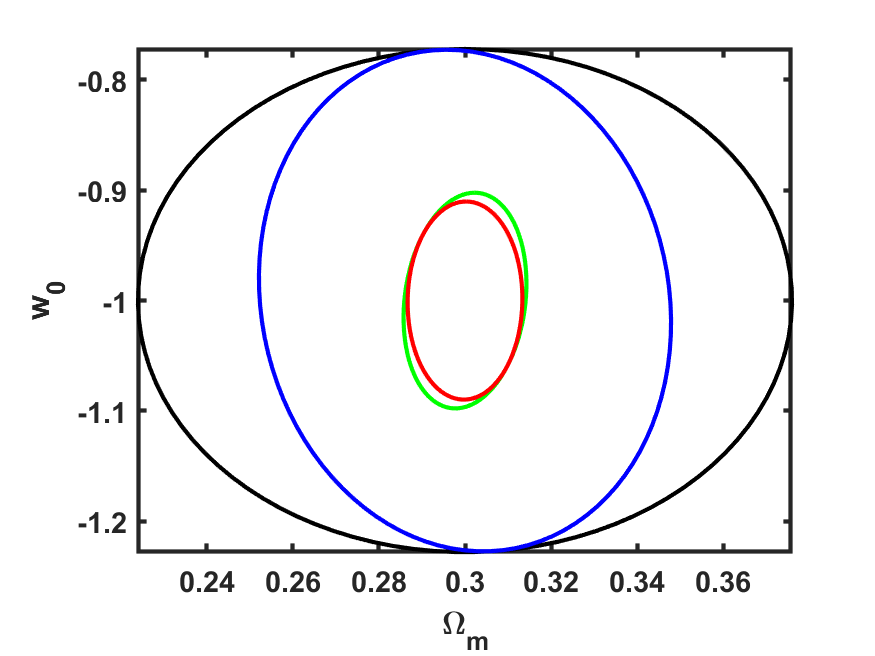}
\includegraphics[width=0.32\textwidth]{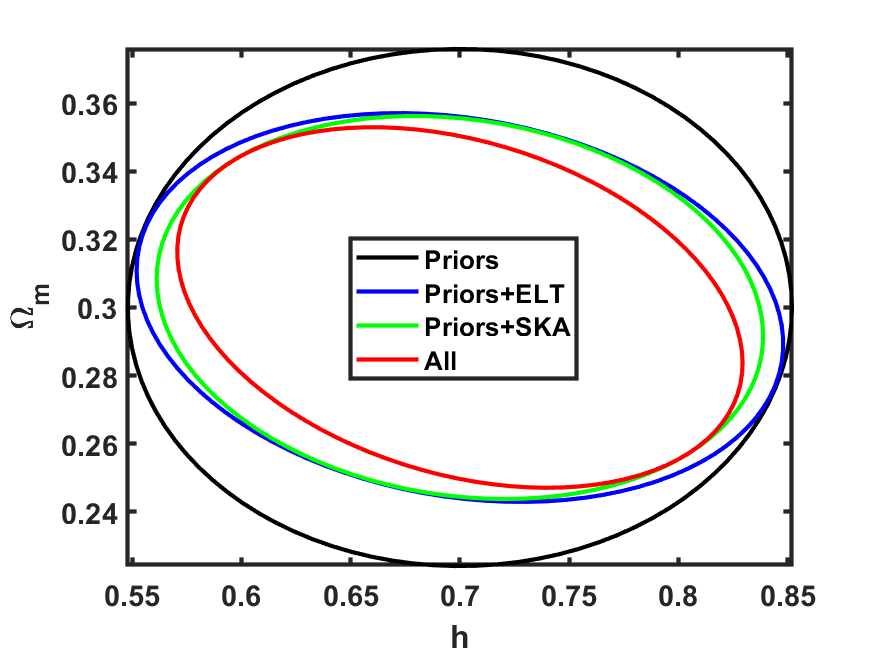}
\includegraphics[width=0.32\textwidth]{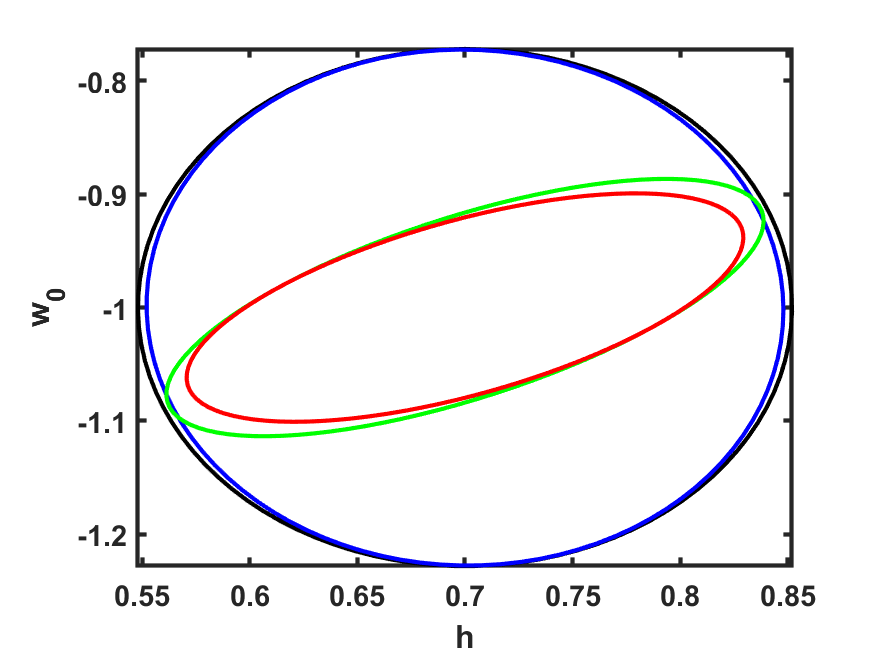}
\includegraphics[width=0.32\textwidth]{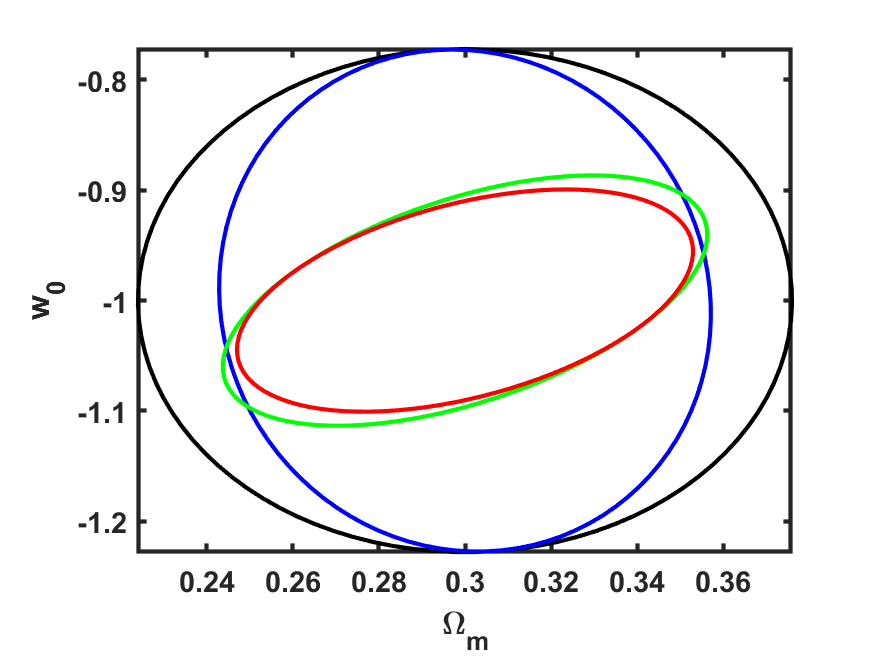}
\includegraphics[width=0.32\textwidth]{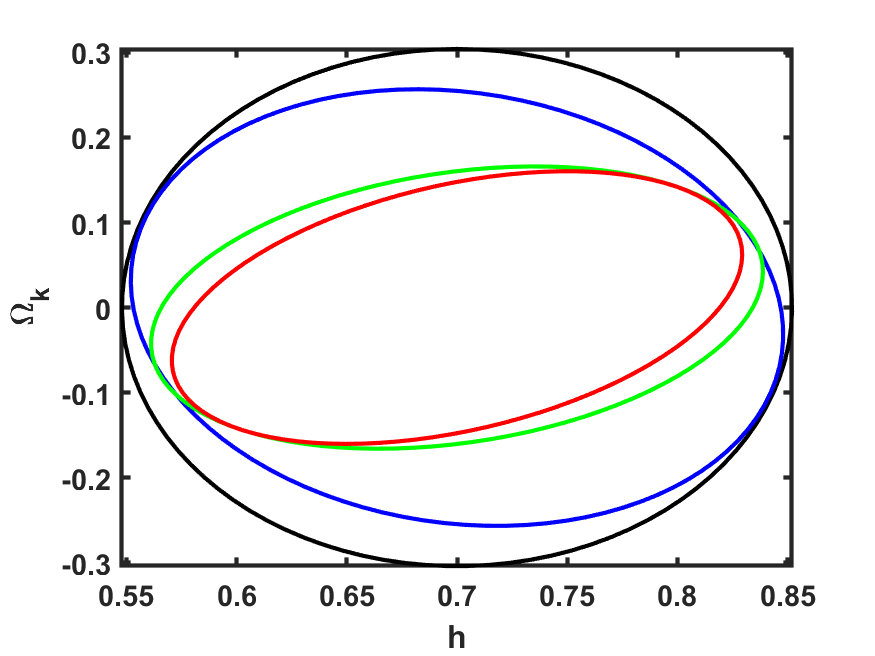}
\includegraphics[width=0.32\textwidth]{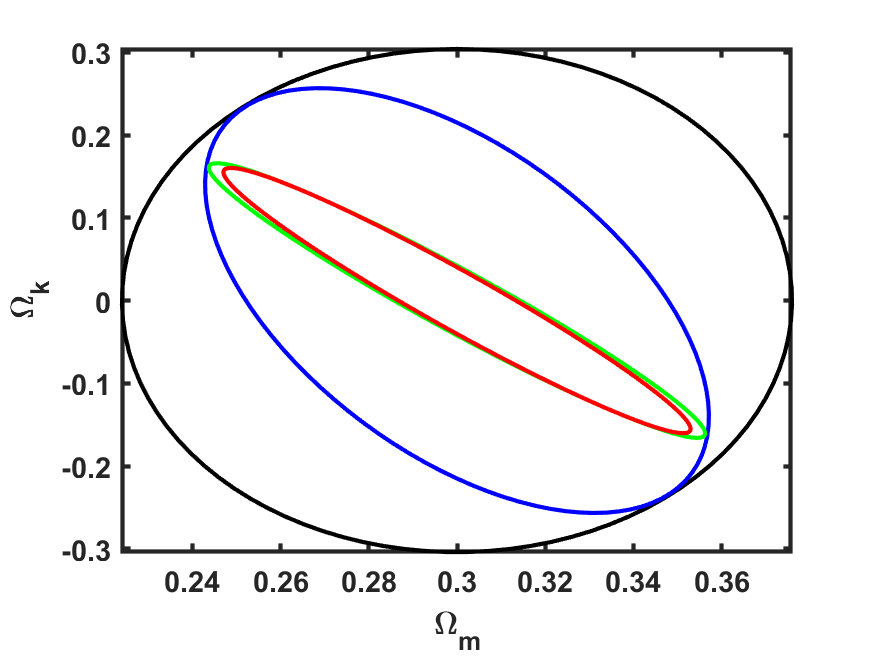}
\includegraphics[width=0.32\textwidth]{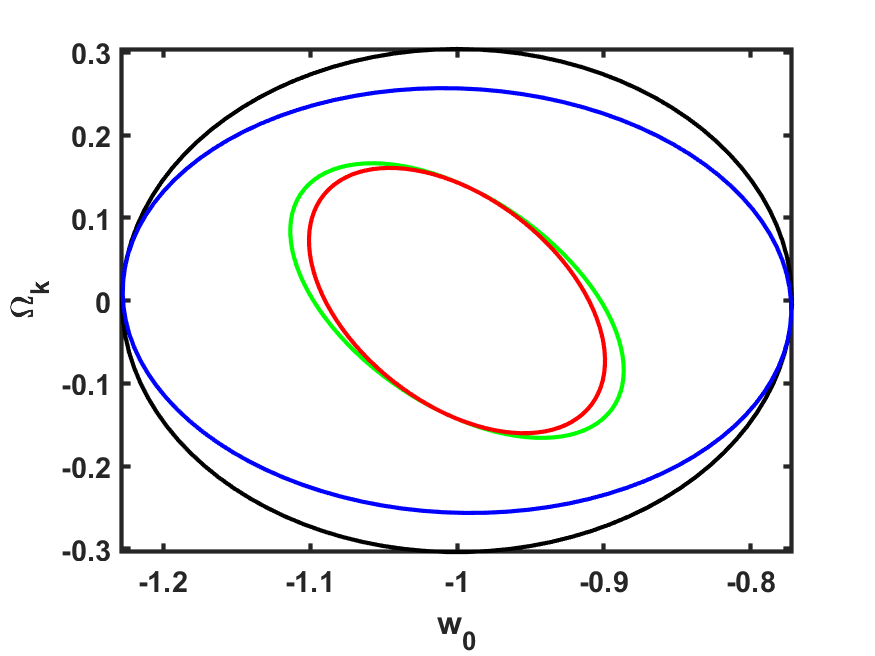}
\end{center}
\caption{Fisher Matrix based forecasts for flat universe fiducial models. In the top row the parameter space does not include $\Omega_k$, while in the middle and bottom rows it does. The black, blue, green and red contours are for Priors only, Priors + ELT, Priors + SKA and Priors +ELT + SKA respectively.}
\label{figure03}
\end{figure*}

\subsection{Flat universes}

In order to provide a baseline for subsequent comparisons, we first consider the case without curvature, specifically with a parameter space $(h,\Omega_m,w_0)$. We assume a fiducial model with $h=0.7$, $\Omega_m=0.3$ and $w_0$, together with the prior fiducial uncertainties (henceforth simply referred to as priors) $\sigma(h)=0.1$, $\sigma(\Omega_m)=0.05$ and $\sigma(w_0)=0.15$. These are meant to represent conservative (i.e. reasonable or slightly pessimistic) current uncertainties on these parameters obtained from contemporary observations.

\begin{table*}
\begin{center}
\caption{Fisher Matrix based forecasts for a fiducial model without a curvature parameter, for various combinations of priors and measurements. The first and second sets of rows show the correlation coefficient and Figure of Merit (FoM) for each pair of parameters, while the final set of rows shows the one-sigma uncertainties for each parameter.}
\label{table1}
\begin{tabular}{c | c c | c c | c}
\hline
Parameter & Priors only & Drift only & SKA + priors & ELT + priors & Drift + priors  \\
\hline
$\rho(h,\Omega_m)$ & 0 & +0.474 & +0.402 & -0.307 & +0.288 \\
$\rho(h,w_0)$ & 0 & +0.990 & +0.965 & -0.015 & +0.961 \\
$\rho(\Omega_m,w_0)$ & 0 & +0.349 & +0.158 & -0.089 & +0.020 \\
\hline
$FoM(h,\Omega_m)$ & 87 & 315 & 573 & 150 & 659 \\
$FoM(h,w_0)$ & 29 & 161 & 294 & 30 & 338 \\
$FoM(\Omega_m,w_0)$ & 58 & 402 & 729 & 93 & 839 \\
\hline
$\sigma(h)$ & 0.100 & 0.161 & 0.088 & 0.097 & 0.079 \\
$\sigma(\Omega_m)$ & 0.050 & 0.010 & 0.009 & 0.032 & 0.009 \\
$\sigma(w_0)$ & 0.150 & 0.119 & 0.064 & 0.149 & 0.059 \\
\hline
\end{tabular}
\end{center}
\end{table*}

The top row of Figure \ref{figure03} shows the results of this analysis, for the priors alone and for the combinations of the priors with the ELT and/or the SKA dataset. Table \ref{table1} provides further details on these results, also including the case of the constraints from the combination of the ELT and SKA measurements without including any priors---in this case, the redshift drift constraints on $\Omega_m$ and $w_0$ are better than the assumed prior (very significantly so in the former case), but this is not the case for $h$. This confirms previous analyses, showing that the redshift drift does not provide stringent constraints on the Hubble parameter, manifestly because it is only a multiplicative factor.

\begin{table*}
\begin{center}
\caption{Same as Table \ref{table1}, but for a fiducial model including the curvature parameter, specifically with the fiducial value $\Omega_k=0$.}
\label{table2}
\begin{tabular}{c | c c | c c | c}
\hline
Parameter & Priors only & Drift only & SKA + priors & ELT + priors & Drift + priors  \\
\hline
$\rho(h,\Omega_m)$ & 0 & -0.977 & -0.149 & -0.189 & -0.309 \\
$\rho(h,w_0)$ & 0 & +0.699 & +0.674 & -0.010 & +0.613 \\
$\rho(\Omega_m,w_0)$ & 0 & -0.535 & +0.526 & -0.052 & +0.443 \\
$\rho(h,\Omega_k)$ & 0 & +0.980 & +0.256 & -0.122 & +0.388 \\
$\rho(\Omega_m,\Omega_k)$ & 0 & -0.999 & -0.967 & -0.546 & -0.968 \\
$\rho(w_0, \Omega_k)$ & 0 & +0.545 & -0.508 & -0.041 & -0.454 \\
\hline
$FoM(h,\Omega_m)$ & 87 & 8 & 130 & 121 & 154 \\
$FoM(h,w_0)$ & 29 & 5 & 86 & 30 & 98 \\
$FoM(\Omega_m,w_0)$ & 58 & 12 & 185 & 77 & 210 \\
$FoM(h,\Omega_k)$ & 22 & 3 & 45 & 27 & 53 \\
$FoM(\Omega_m,\Omega_k)$ & 43 & 45 & 424 & 82 & 470 \\
$FoM(w_0, \Omega_k)$ & 14 & 4 & 62 & 17 & 70 \\
\hline
$\sigma(h)$ & 0.100 & 0.806 & 0.091 & 0.097 & 0.085 \\
$\sigma(\Omega_m)$ & 0.050 & 0.314 & 0.037 & 0.038 & 0.035 \\
$\sigma(w_0)$ & 0.150 & 0.142 & 0.075 & 0.149 & 0.066 \\
$\sigma(\Omega_k)$ & 0.200 & 1.001 & 0.109 & 0.169 & 0.106 \\
\hline
\end{tabular}
\end{center}
\end{table*}

We can now repeat the above analysis, still for a flat fiducial universe ($\Omega_k=0$) but including curvature in the parameter space, which becomes $(h,\Omega_m,w_0,\Omega_k)$. For the curvature parameter we assume a prior uncertainty $\sigma(\Omega_k)=0.2$. This is a more conservative prior than those for the other model parameters, which is justified since the study of the impact of (and sensitivity to) this parameter is the focus of the present work. The middle and bottom rows of Figure \ref{figure03} and also Table \ref{table2} summarize the results.

As expected, the constraints become weaker, with this weakening being more noticeable for the SKA, and especially for the matter density. The latter aspect is also unsurprising given the strong anti-correlation between $\Omega_m$ and $\Omega_k$. Nevertheless, it is interesting to note that the assumed combination of SKA and ELT measurements of the redshift drift can, even in the absence of any priors, provide a non-trivial constraint on the dark energy equation of state, $w_0$, which is better than our assumed prior for this parameter.

\subsection{Open and closed universes}

\begin{figure*}
\begin{center}
\includegraphics[width=0.32\textwidth]{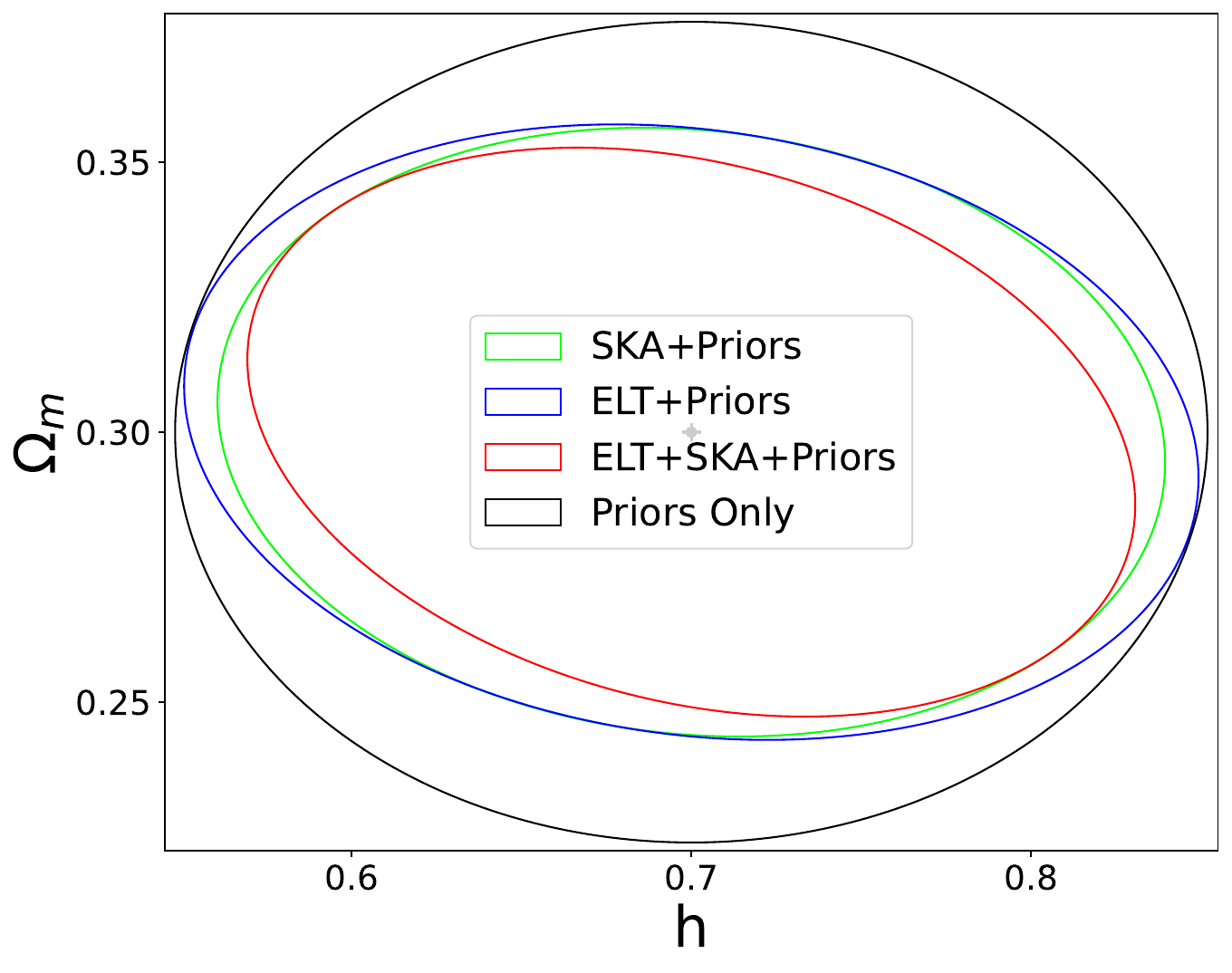}
\includegraphics[width=0.32\textwidth]{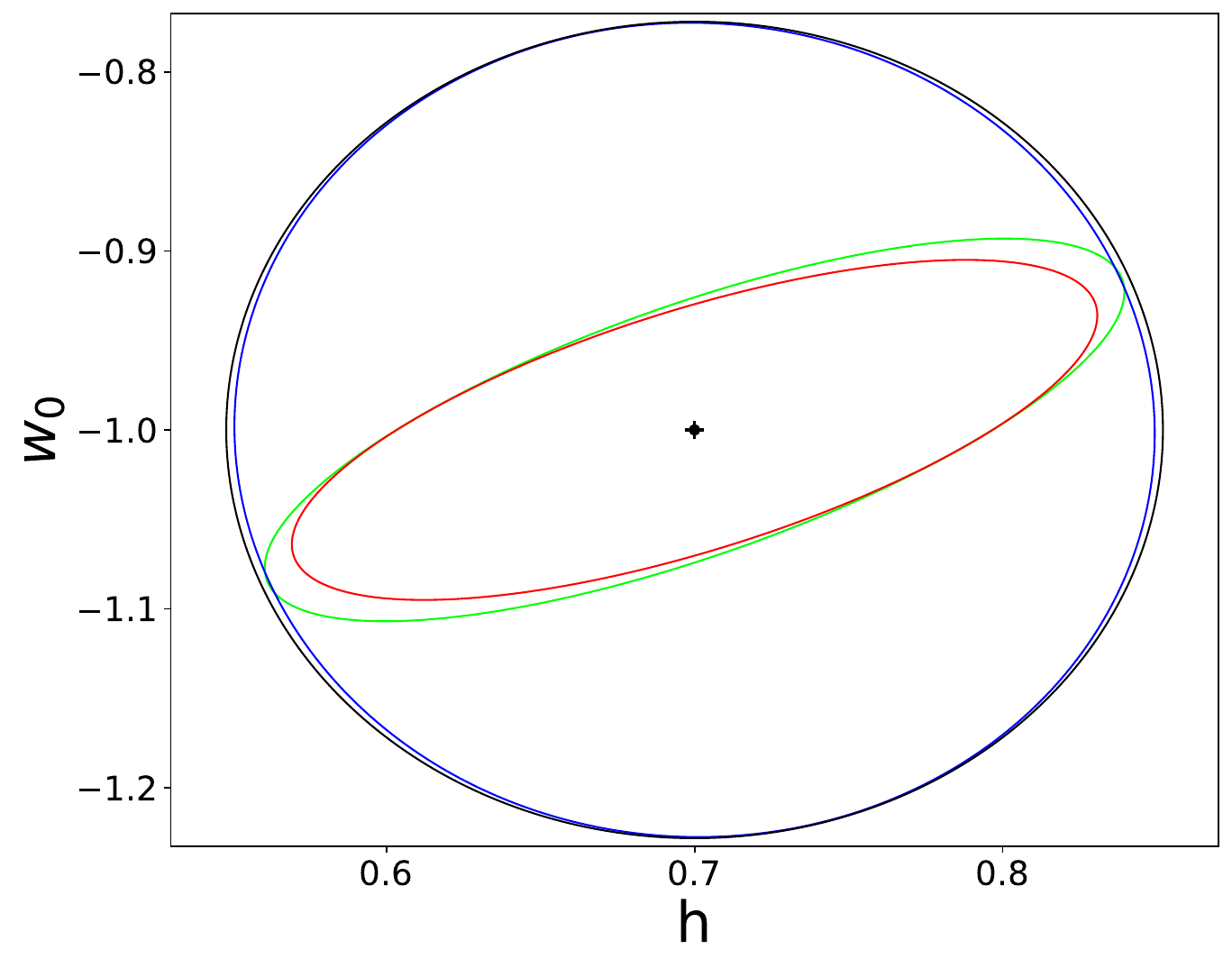}
\includegraphics[width=0.32\textwidth]{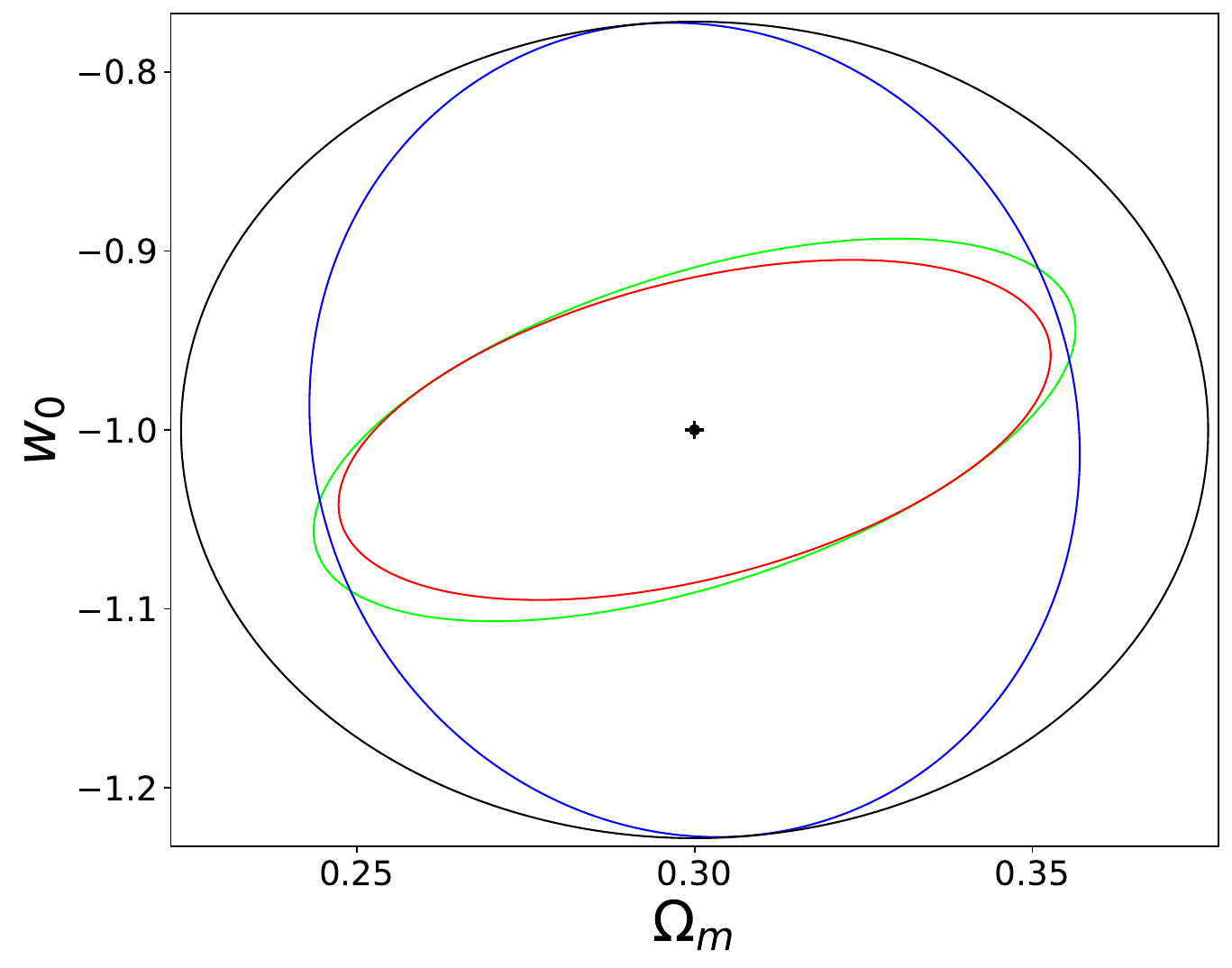}
\includegraphics[width=0.32\textwidth]{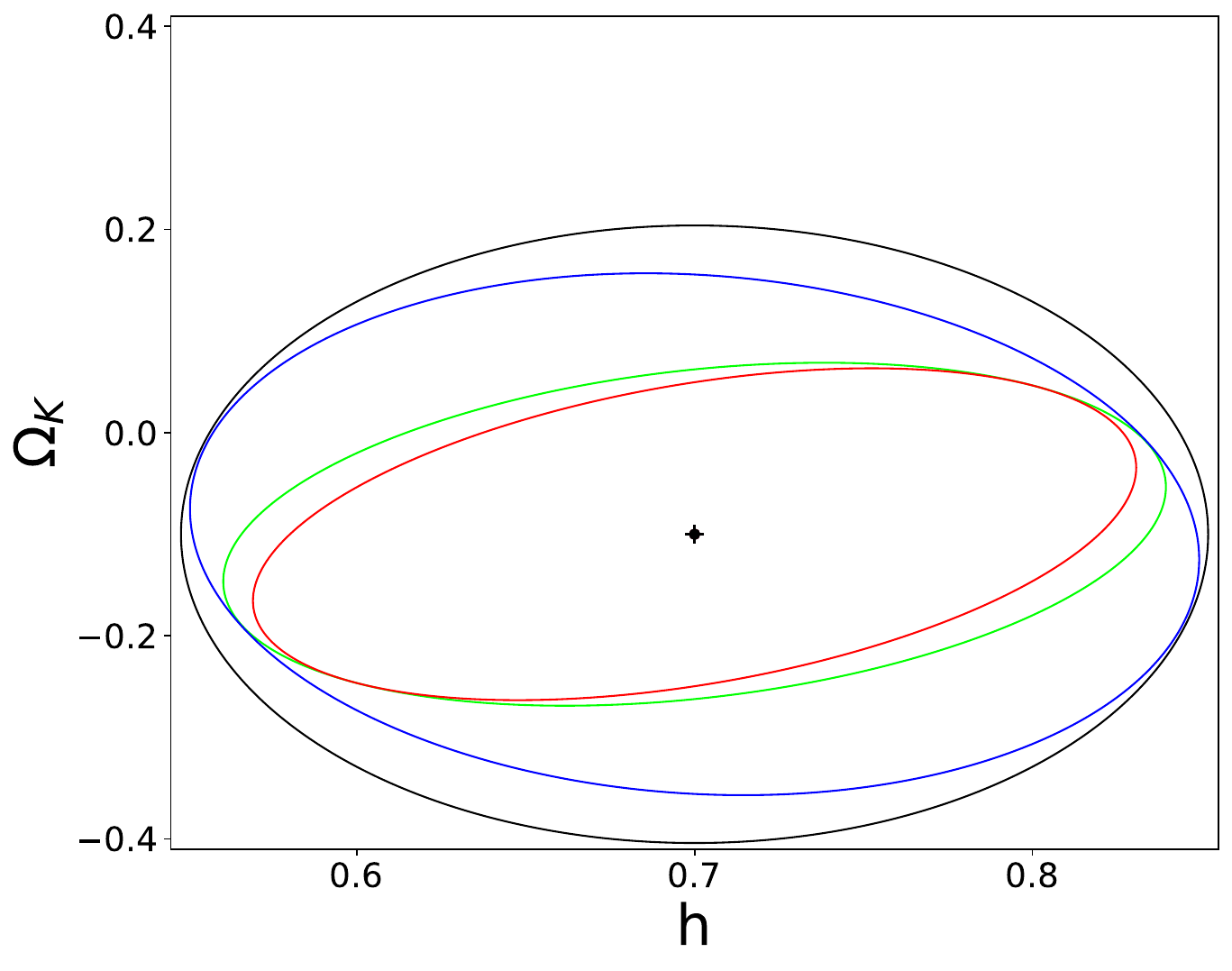}
\includegraphics[width=0.32\textwidth]{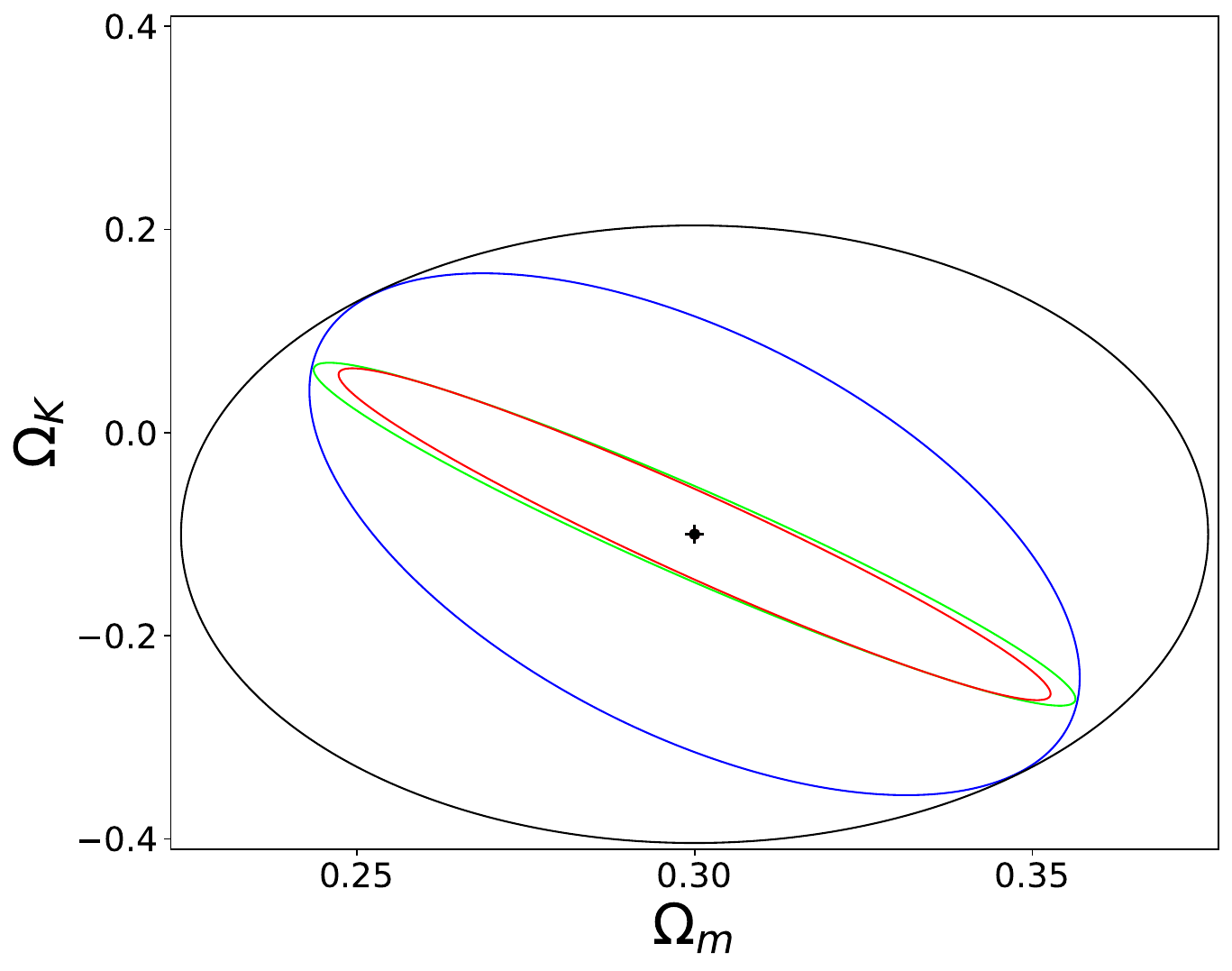}
\includegraphics[width=0.32\textwidth]{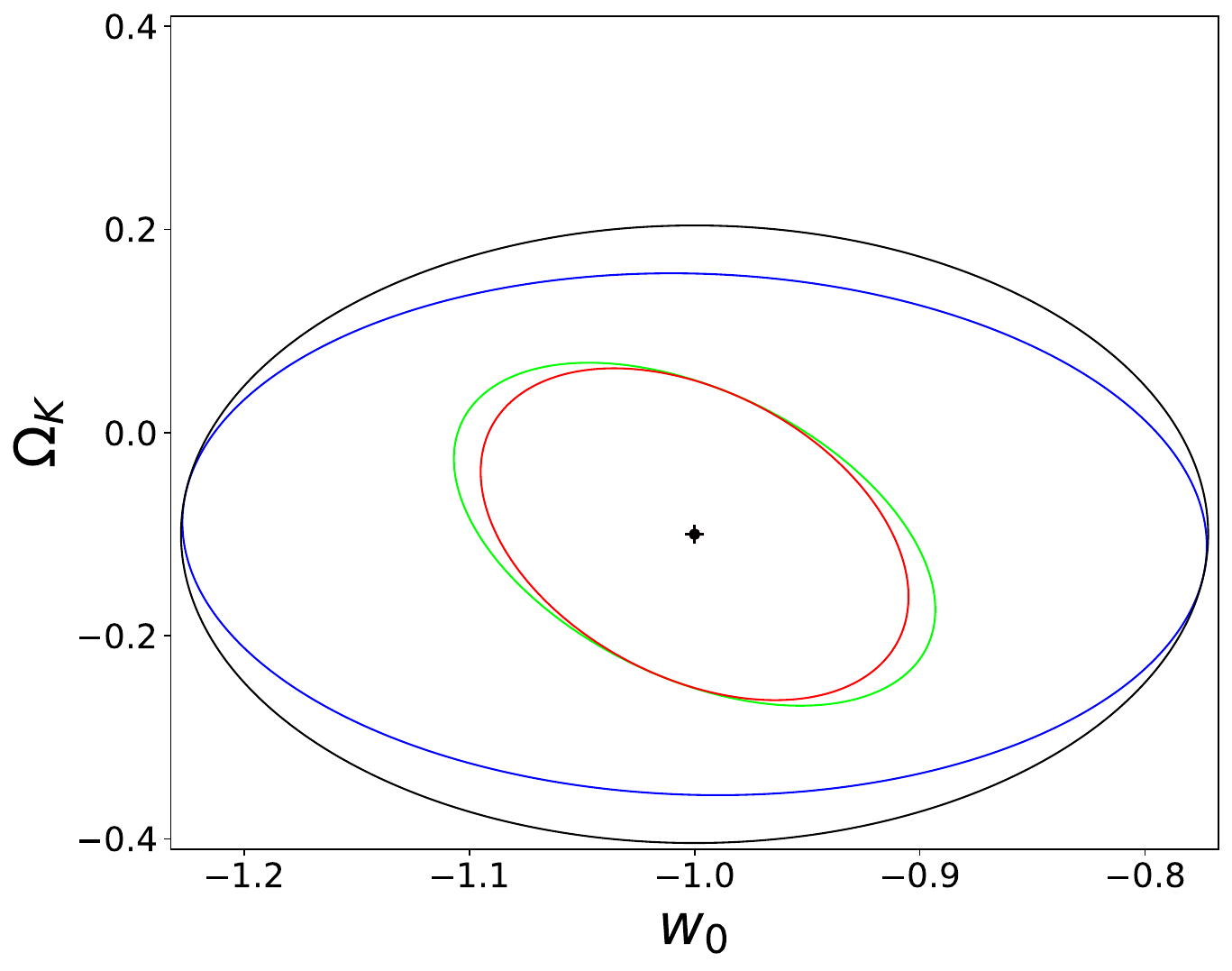}
\includegraphics[width=0.32\textwidth]{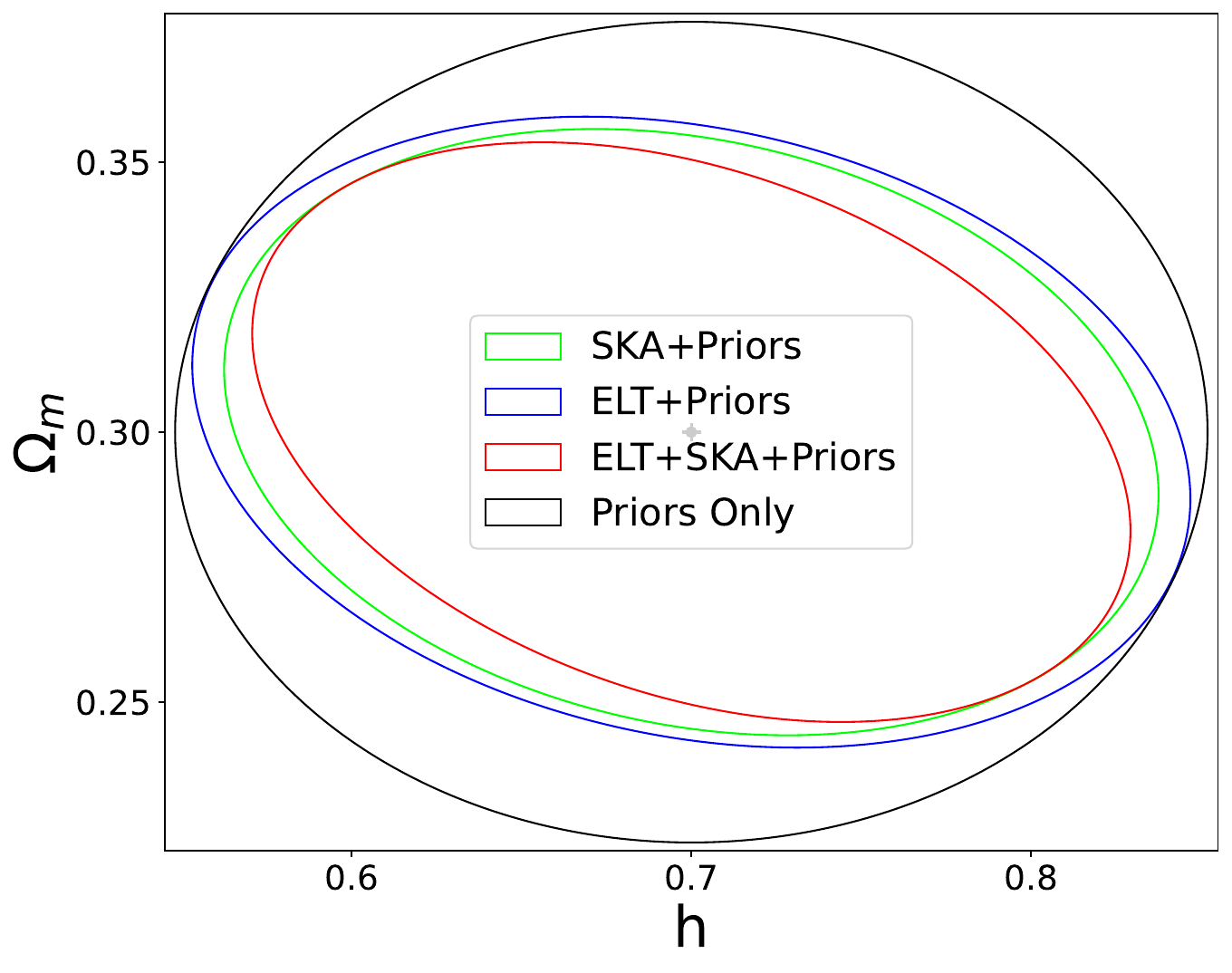}
\includegraphics[width=0.32\textwidth]{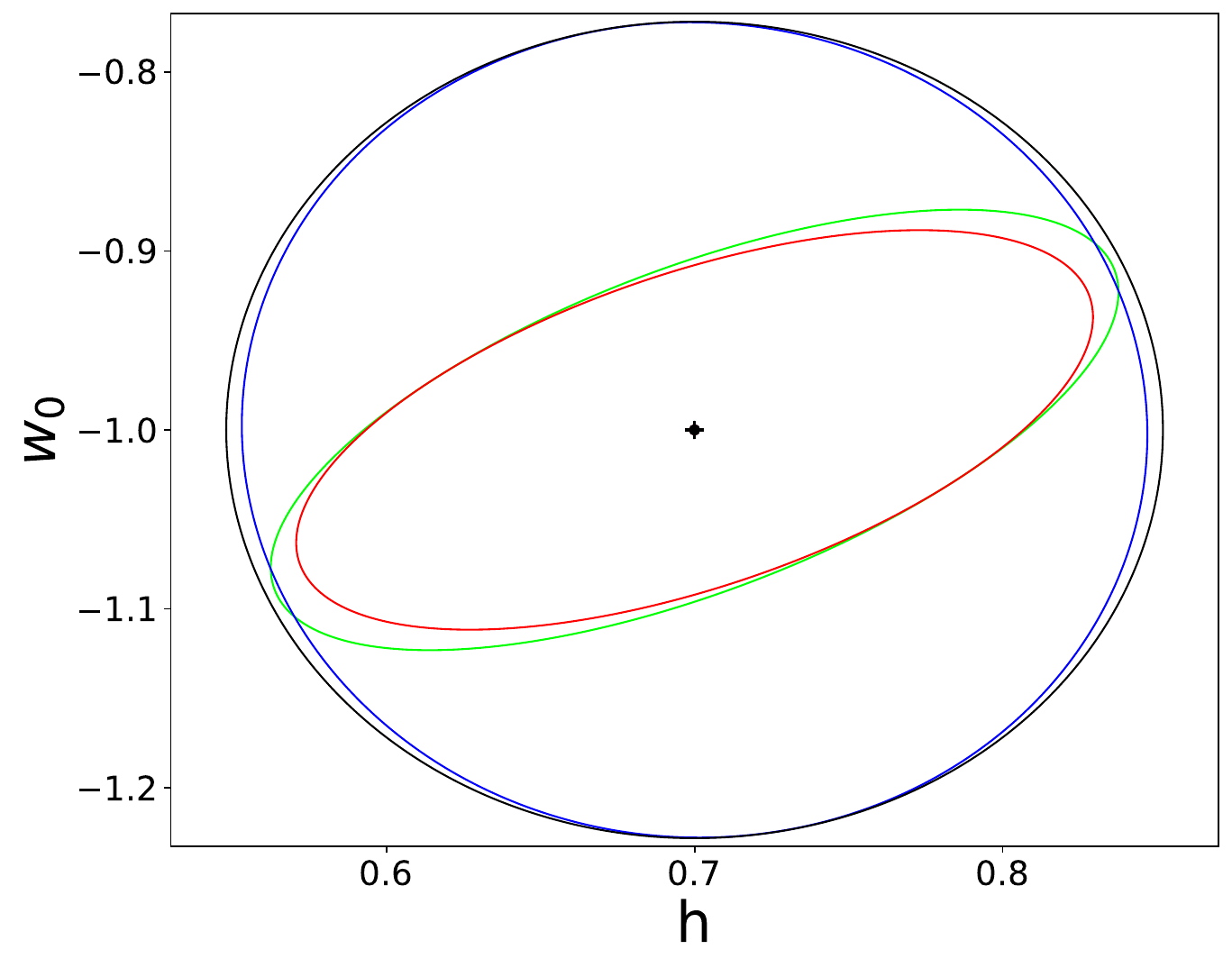}
\includegraphics[width=0.32\textwidth]{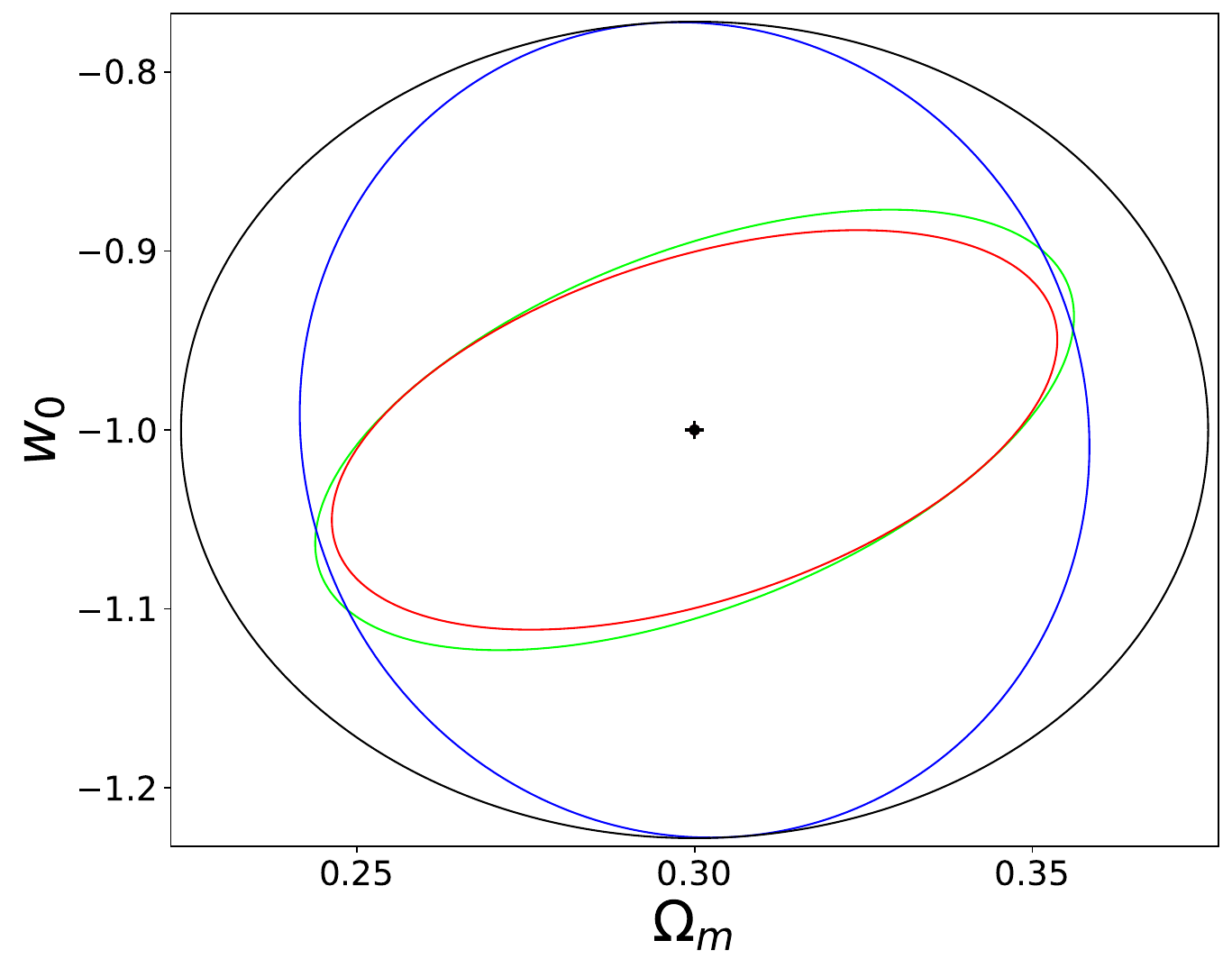}
\includegraphics[width=0.32\textwidth]{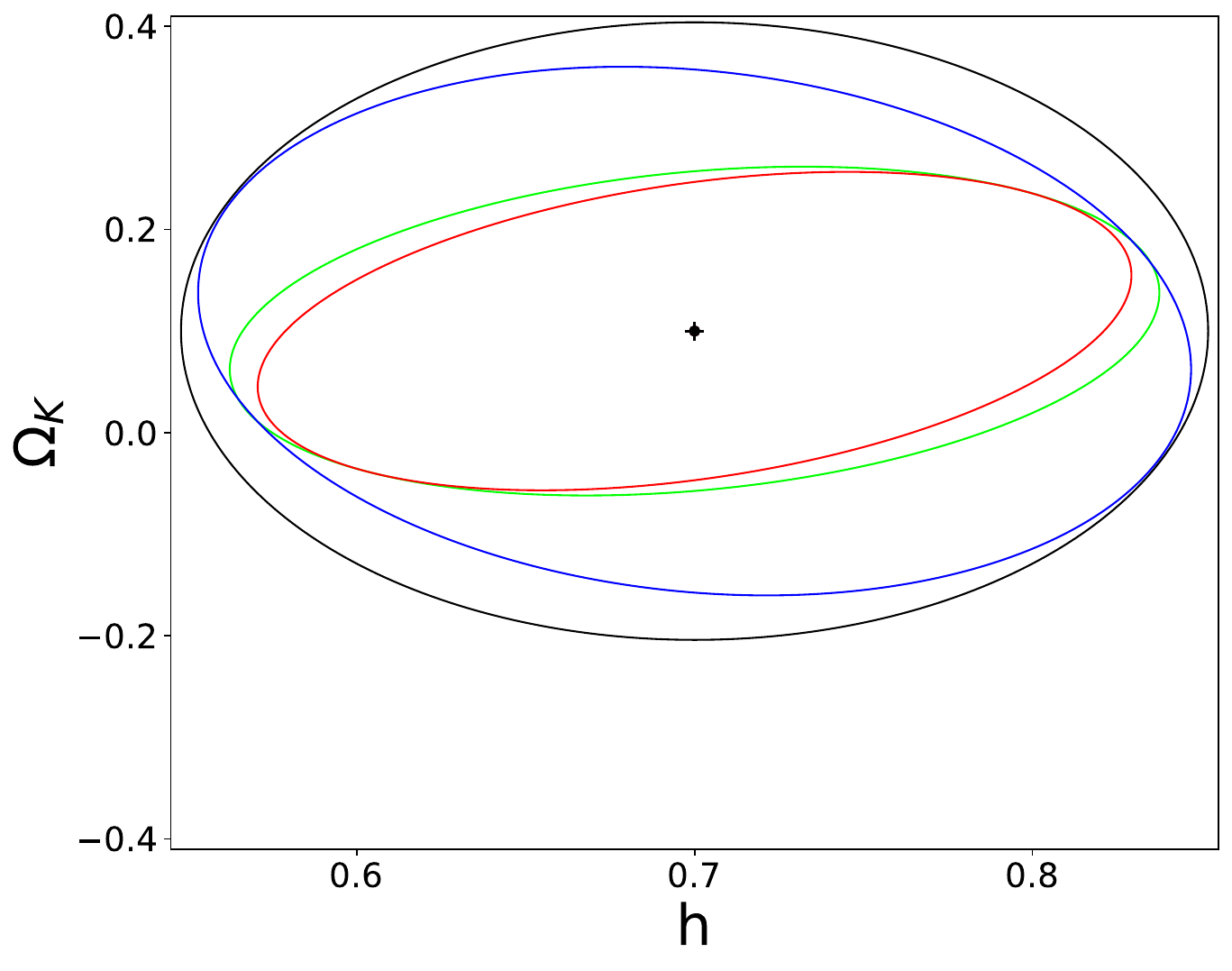}
\includegraphics[width=0.32\textwidth]{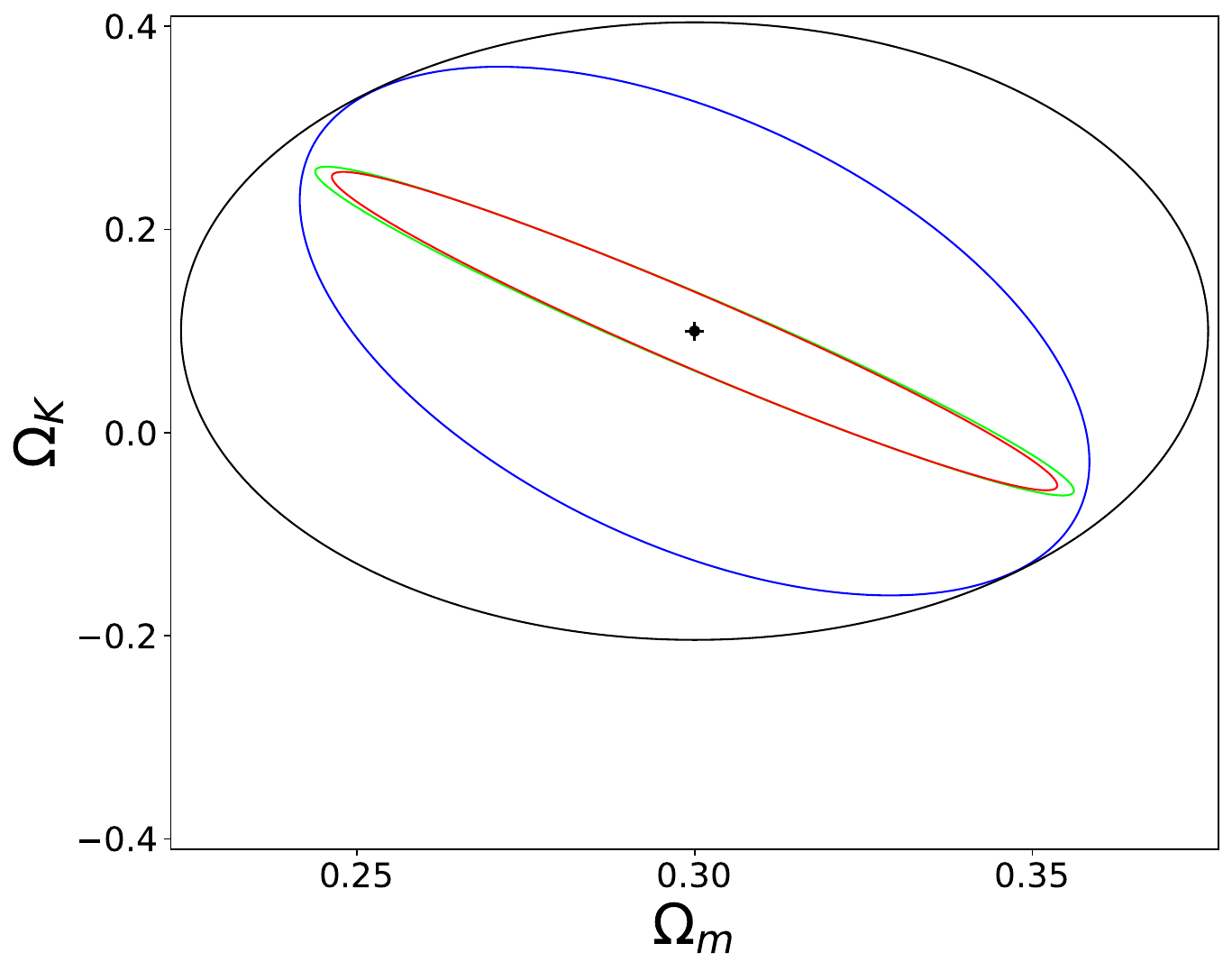}
\includegraphics[width=0.32\textwidth]{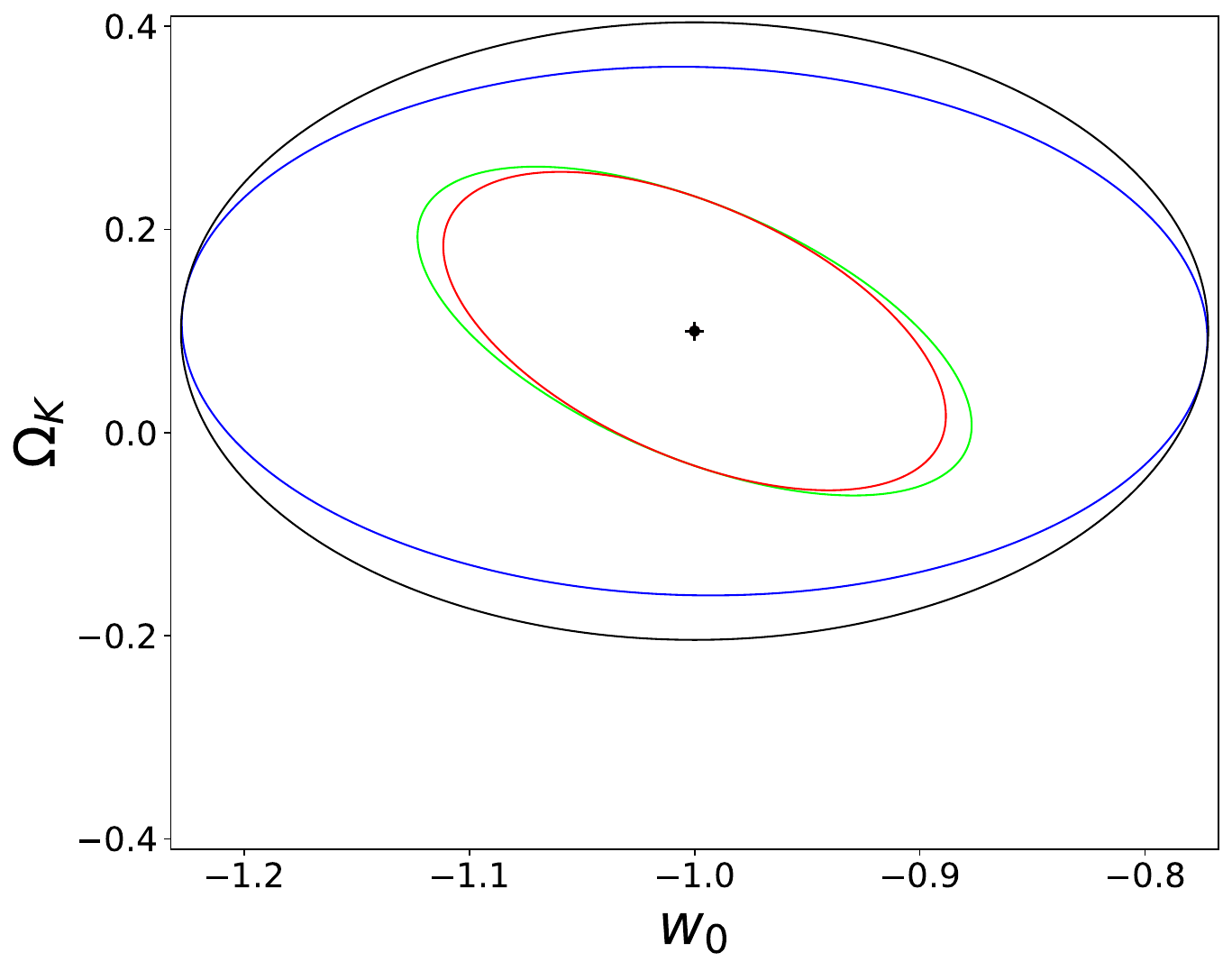}
\end{center}
\caption{Fisher Matrix based forecasts for closed and open universes (first and second, and third and fourth rows, respectively). The solid, dashed and dotted lines correspond to $\Omega_k=0$, $\Omega_k=-0.1$ and $\Omega_k=+0.1$ respectively. The black, blue, green and red contours are for Priors only, Priors + ELT, Priors + SKA and Priors +ELT + SKA respectively. To facilitate the comparison, corresponding panels have identical axis ranges in both cases.}
\label{figure04}
\end{figure*}

We now repeat the analysis in the previous subsection under the assumption of open and closed universes, specifically for fiducial values of the curvature parameter $\Omega_k=\pm0.1$. All the other independent parameters remain the same as in the previous subsection. The results of this analysis are summarized in Figure \ref{figure04} (to be compared to the middle and bottom rows of Figure \ref{figure03}, which contain the flat case) and in Tables \ref{table3}--\ref{table4}---which can be compared to Table \ref{table2}.

\begin{table*}
\begin{center}
\caption{Same as Table \ref{table2}, but for a fiducial open universe model, specifically $\Omega_k=+0.1$.}
\label{table3}
\begin{tabular}{c | c c | c c | c}
\hline
Parameter & Priors only & Drift only & SKA + priors & ELT + priors & Drift + priors  \\
\hline
$\rho(h,\Omega_m)$ & 0 & -0.981 & -0.255 & -0.211 & -0.370 \\
$\rho(h,w_0)$ & 0 & +0.552 & +0.634 & -0.011 & +0.579 \\
$\rho(\Omega_m,w_0)$ & 0 & -0.385 & +0.517 & -0.041 & +0.465 \\
$\rho(h,\Omega_k)$ & 0 & +0.981 & +0.268 & -0.144 & +0.374 \\
$\rho(\Omega_m,\Omega_k)$ & 0 & -0.999 & -0.552 & -0.495 & -0.986 \\
$\rho(w_0, \Omega_k)$ & 0 & +0.382 & -0.513 & -0.031 & -0.513 \\
\hline
$FoM(h,\Omega_m)$ & 87 & 8 & 140 & 120 & 160 \\
$FoM(h,w_0)$ & 29 & 5 & 81 & 30 & 90 \\
$FoM(\Omega_m,w_0)$ & 58 & 11 & 183 & 76 & 203 \\
$FoM(h,\Omega_k)$ & 22 & 4 & 48 & 27 & 56 \\
$FoM(\Omega_m,\Omega_k)$ & 43 & 69 & 648 & 76 & 675 \\
$FoM(w_0, \Omega_k)$ & 14 & 4 & 65 & 17 & 72 \\
\hline
$\sigma(h)$ & 0.100 & 0.839 & 0.089 & 0.097 & 0.085 \\
$\sigma(\Omega_m)$ & 0.050 & 0.333 & 0.036 & 0.038 & 0.034  \\
$\sigma(w_0)$ & 0.150 & 0.132 & 0.077 & 0.149 & 0.069 \\
$\sigma(\Omega_k)$ & 0.200  & 0.976 & 0.104 & 0.171 & 0.101 \\
\hline
\end{tabular}
\end{center}
\end{table*}
\begin{table*}
\begin{center}
\caption{Same as Table \ref{table2}, but for a fiducial closed universe model, specifically $\Omega_k=-0.1$.}
\label{table4}
\begin{tabular}{c | c c | c c | c}
\hline
Parameter & Priors only & Drift only & SKA + priors & ELT + priors & Drift + priors  \\
\hline
$\rho(h,\Omega_m)$ & 0 & -0.969 & -0.079 & -0.149 & -0.246 \\
$\rho(h,w_0)$ & 0 & +0.836 & +0.705 & -0.009 & +0.653 \\
$\rho(\Omega_m,w_0)$ & 0 & -0.684 & +0.513 & -0.045 & +0.419 \\
$\rho(h,\Omega_k)$ & 0 & +0.977 & +0.261 & -0.144 & +0.389 \\
$\rho(\Omega_m,\Omega_k)$ & 0 & -0.999 & -0.944 & -0.550 & -0.947 \\
$\rho(w_0, \Omega_k)$ & 0 &  +0.704 & +0.457 & -0.096 & -0.399 \\
\hline
$FoM(h,\Omega_m)$ & 87 & 7 & 125 & 119 & 148 \\
$FoM(h,w_0)$ & 29 & 6 & 92 & 30 & 103 \\
$FoM(\Omega_m,w_0)$ & 58 & 11 & 185 & 75 & 210 \\
$FoM(h,\Omega_k)$ & 22 & 2 & 43 & 26 & 50 \\
$FoM(\Omega_m,\Omega_k)$ & 43 & 30 & 310 & 82 & 356 \\
$FoM(w_0, \Omega_k)$ & 14  & 3 & 60 & 17 & 67 \\
\hline
$\sigma(h)$ & 0.100 &  0.809 & 0.092 & 0.098 & 0.086 \\
$\sigma(\Omega_m)$ & 0.050 & 0.297 & 0.038 & 0.037 & 0.035 \\
$\sigma(w_0)$ & 0.150 & 0.175 & 0.072 & 0.149 & 0.065 \\
$\sigma(\Omega_k)$ & 0.200 & 1.020 & 0.113 & 0.169 & 0.109 \\
\hline
\end{tabular}
\end{center}
\end{table*}

We find that for most parameters, the impact of curvature on the derived one-sigma constraints $\sigma_i$ is small. This is particularly the case for the ELT, whose sensitivity to curvature is smaller than that of the SKA. If considering redshift drift measurements alone (without invoking any priors), the only Figure of Merit (FoM) which changes substantially is the one in the $(\Omega_m,\Omega_k)$ plane, which changes by more than a factor of two between closed and open universes (being larger in the latter case). This is to be expected, since the matter, curvature and dark energy terms in the Friedmann equation have different redshift dependencies. By changing $\Omega_k$ one changes their relative weights (since the sum of the $\Omega_i$ is unity). Since the redshift at which the measurements are made is (by assumption) unchanged, the sensitivity to the parameters changes and so does the FoM.

More generally, whether or not priors are included, the constraints on the dark energy equation of state $w_0$ are slightly better in closed universes, while for the other parameters the differences are smaller. These results are consistent with the behaviour of the sensitivity coefficient discussed in Section  \ref{stddri}. Overall, the effect of extending the parameter space by including $\Omega_k$ is much larger than the differences between the open, flat, and closed cases.

\section{Differential redshift drift}
\label{difdri}

An alternative approach to high-resolution spectroscopy redshift drift measurements, relying on the Lyman-$\alpha$ forest, proposes to measure the drift of the relative redshift between two astrophysical sources along the same line of sight \cite{Cooke}. In what follows, for simplicity we will refer to this as the differential redshift drift. In this section we provide a discussion of the cosmological sensitivity of such measurements.

A possible methodological advantage of this approach is being less reliant on the stringent wavelength calibration requirements than the standard approach, as well as being less vulnerable to barycentric correction uncertainties---although these need to be quantified by detailed simulations. On the other hand, this technique would only be applicable to a small number of wavelength regions of the observed spectrum, not to the full Lyman-$\alpha$ forest as in the standard technique, which would impact the available signal-to-noise for the same amount of observing time. Strictly speaking, this approach is only applicable to the ELT ANDES measurements (not to the SKA), but in this section we take a conceptual approach and explore the potential impact of such measurements at arbitrary redshifts. We keep the same assumptions on fiducial models as in Section \ref{stddri}.

For a reference redshift $z_r$ and an intervening redshift $z_i<z_r$, we define\footnote{Note that in what follows our definitions differ, by a minus sign, from those of \cite{Cooke,Esteves}. This change is made with the goal of exactly recovering the standard redshift drift when taking $z_i=0$.}
\be
\frac{\Delta z_{ri}}{\Delta t}=\frac{\Delta z_{r}}{\Delta t}-\frac{\Delta z_{i}}{\Delta t}\,,
\ee
which in dimensionless form becomes
\be
S_z(z_r,z_i)=h \left[(z_r-z_i)- (E(z_r)-E(z_i))\right]\,.
\ee
Similarly, for the spectroscopic velocity we have
\be\label{difdeltav}
S_v(z_r,z_i)=kh\left[ \frac{E(z_i)}{1+z_i} -  \frac{E(z_r)}{1+z_r}  \right]\,.
\ee

\begin{figure*}
\begin{center}
\includegraphics[width=0.49\textwidth]{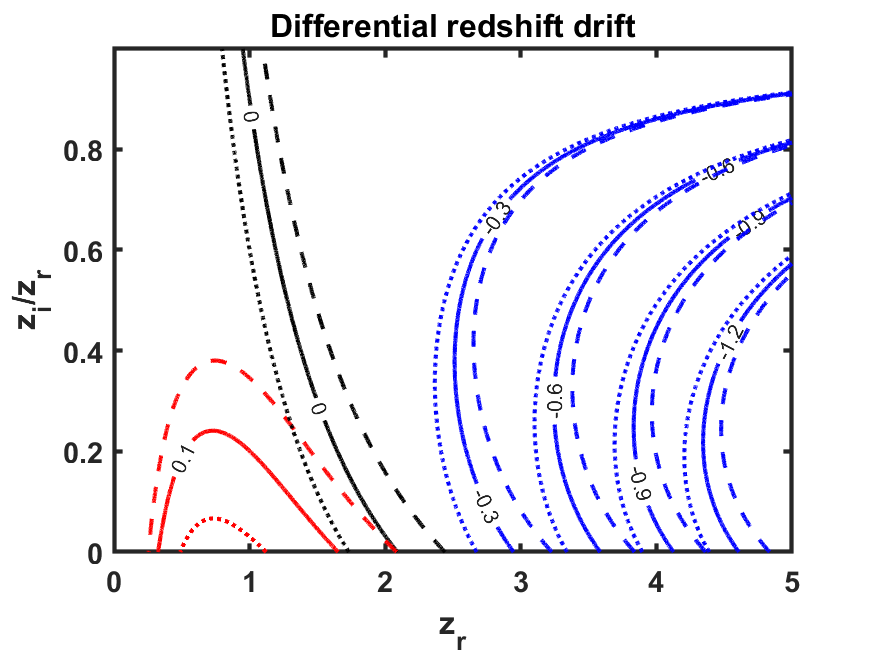}
\includegraphics[width=0.49\textwidth]{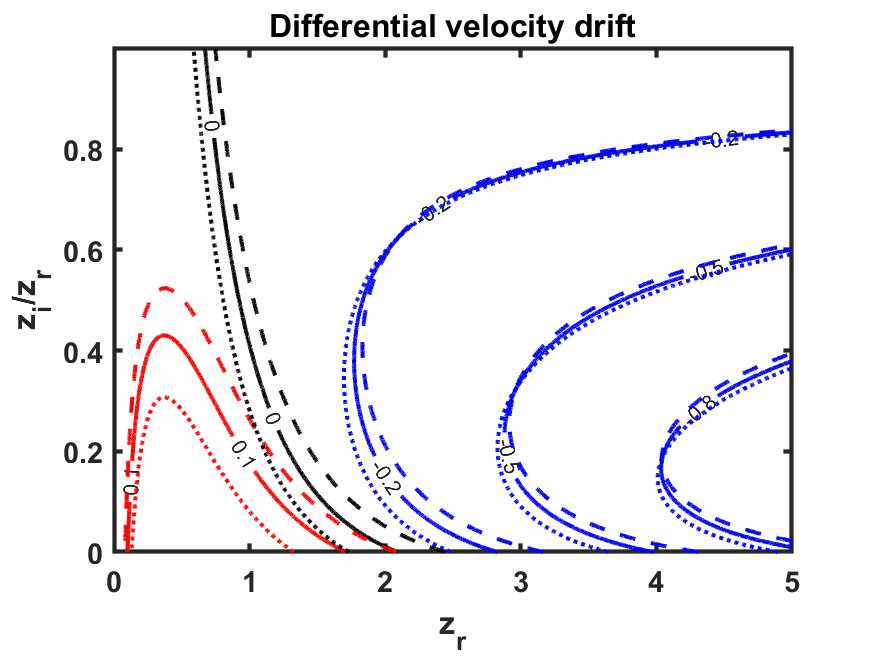}
\end{center}
\caption{Illustrating the behaviour of the differential redshift drift (left panel, in dimensionless units) and the corresponding spectroscopic velocity (right panel, in units of cm/s) for three different fiducial models. Negative, zero and positive valued contours are shown in blue, black, and red respectively. The solid, dashed, and dotted lines correspond to $\Omega_k=0$, $\Omega_k=-0.1$, and $\Omega_k=+0.1$ respectively.}
\label{figure05}
\end{figure*}

Since we now have a two-parameter function, its visualization is somewhat less obvious, but Figure \ref{figure05} illustrates the overall behaviour, depicting selected contour lines for the values of the differential redshift drift and the corresponding spectroscopic velocity for three different (open, flat, and closed) fiducial models. The reference redshift $z_r$ is plotted in the horizontal axis, while the vertical axis shows the ratio $z_i/z_r$, which always spans the range between zero and unity. It should be clear that when this ratio is zero one recovers the behaviour of the standard redshift drift.

\begin{figure*}
\begin{center}
\includegraphics[width=0.49\textwidth]{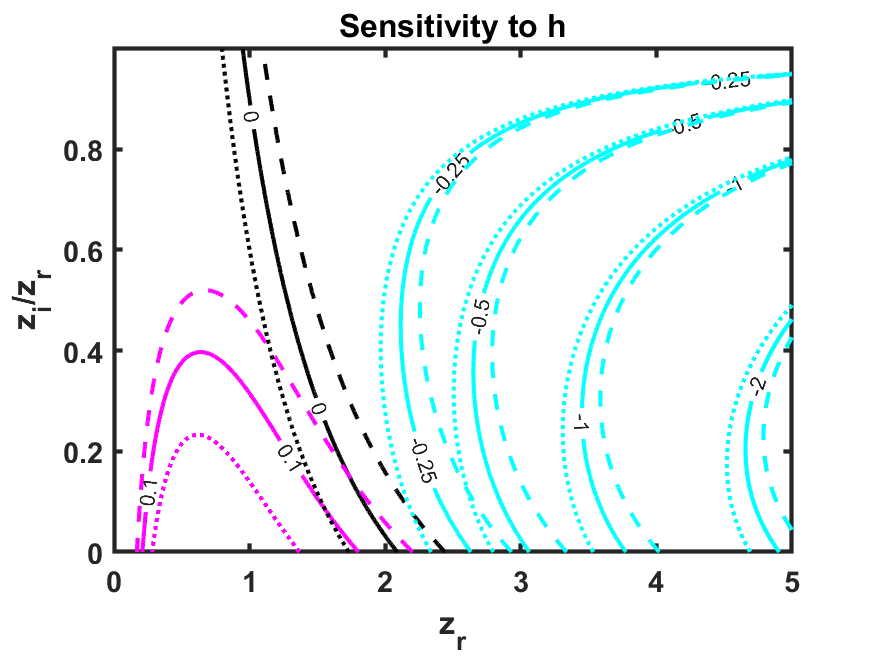}
\includegraphics[width=0.49\textwidth]{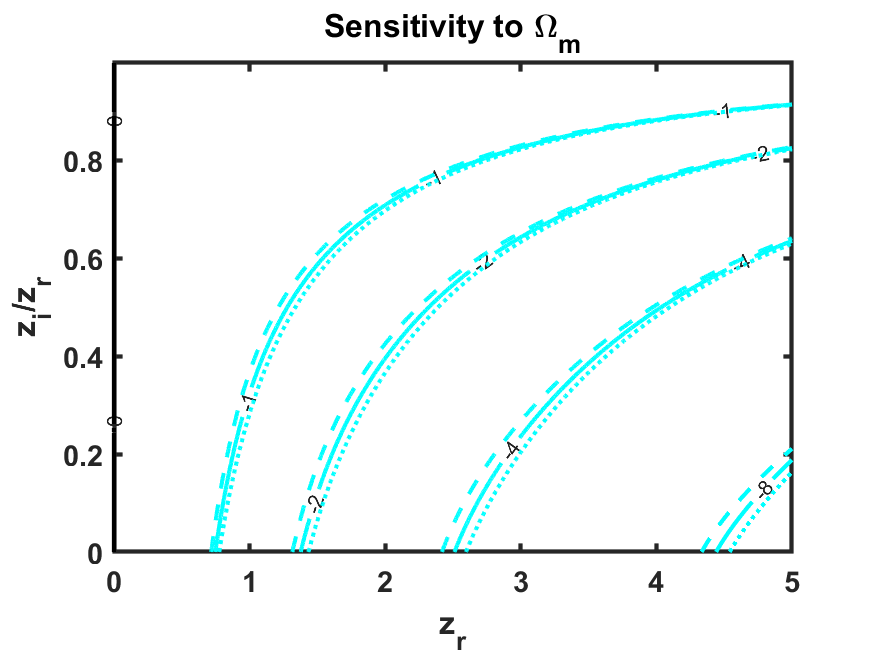}
\includegraphics[width=0.49\textwidth]{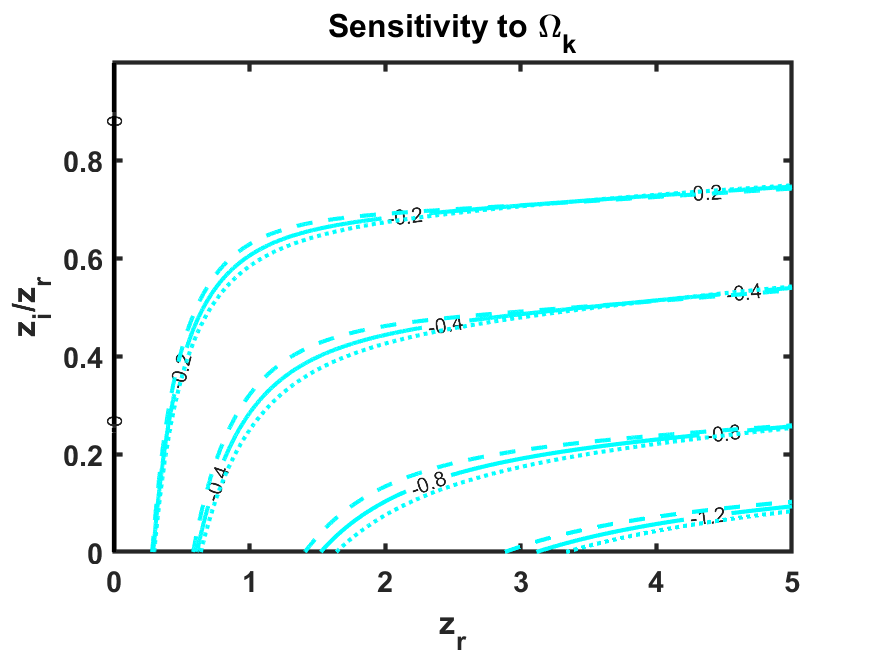}
\includegraphics[width=0.49\textwidth]{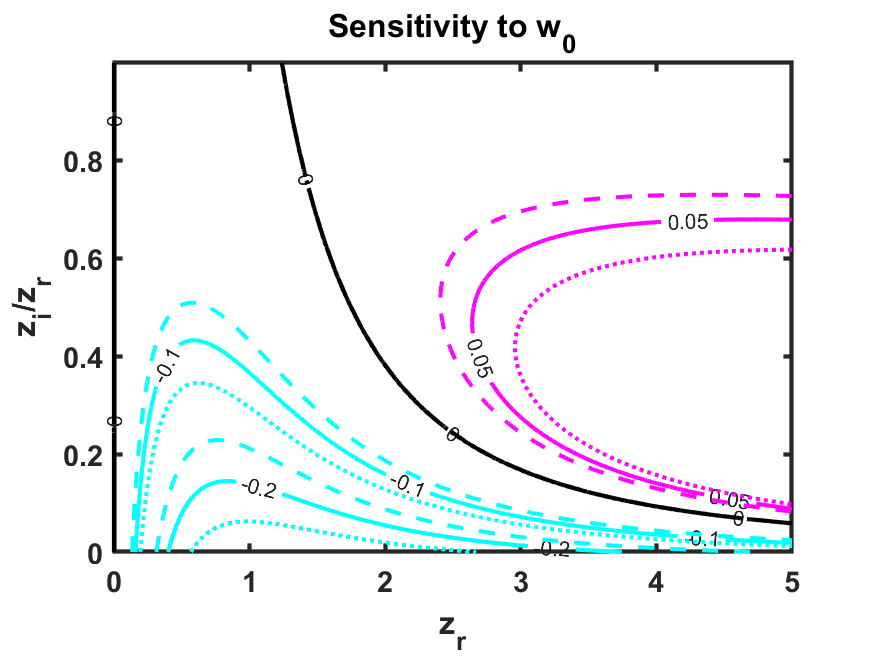}
\end{center}
\caption{Theoretical sensitivity coefficients (in dimensionless units) of the differential redshift drift, for three different fiducial models, to the cosmological parameters: $h$ (top left), $\Omega_m$ (top right), $\Omega_k$ (bottom left) and $w_0$ (bottom right). Negative, zero, and positive valued contours are shown in cyan, black, and magenta respectively. The solid, dashed and dotted lines correspond to $\Omega_k=0$, $\Omega_k=-0.1$, and $\Omega_k=+0.1$ respectively.}
\label{figure06}
\end{figure*}

\begin{figure*}
\begin{center}
\includegraphics[width=0.49\textwidth]{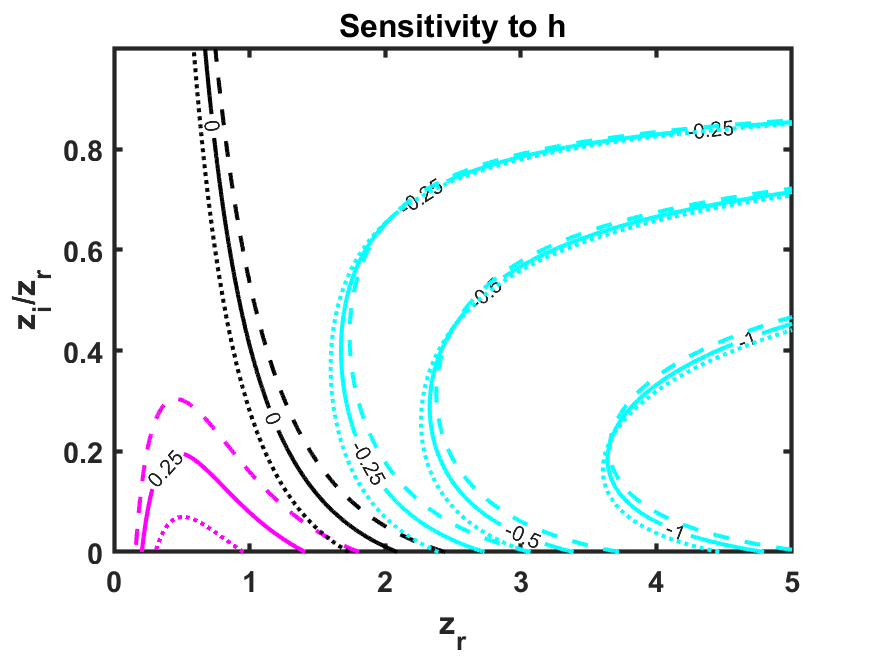}
\includegraphics[width=0.49\textwidth]{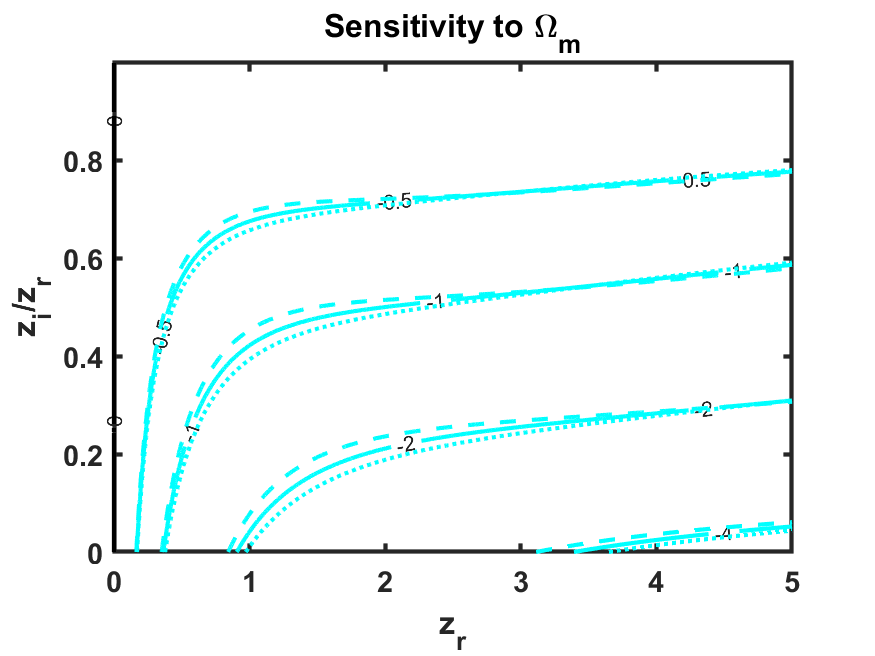}
\includegraphics[width=0.49\textwidth]{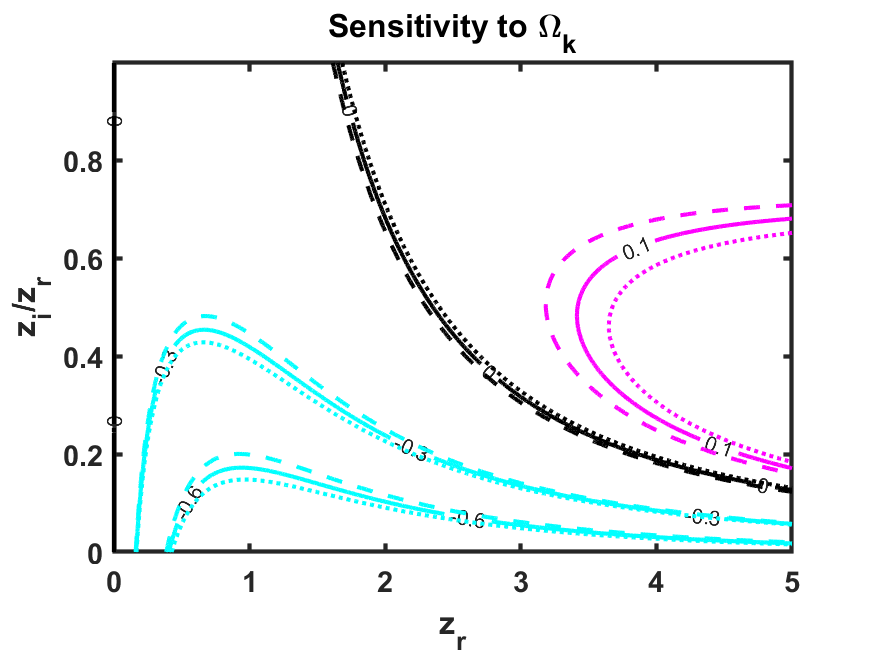}
\includegraphics[width=0.49\textwidth]{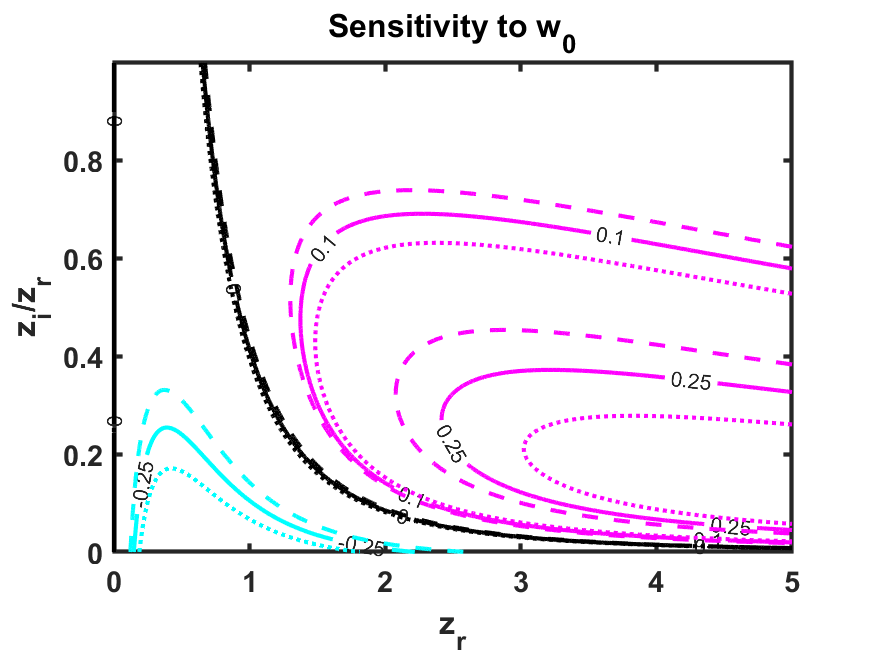}
\end{center}
\caption{Same as Figure \ref{figure06} for the spectroscopic velocity.}
\label{figure07}
\end{figure*}

Along the same lines, Figures \ref{figure06} and \ref{figure07} depict the sensitivity coefficients of the differential redshift drift and the corresponding spectroscopic velocity to the four model parameters. It is worthy of note that, just like in the more standard case, for the differential redshift drift itself the behaviour of the $\Omega_k$ sensitivity is qualitatively similar to that of $\Omega_m$, while in terms of the corresponding spectroscopic velocity it is qualitatively similar to that of $w_0$.

\begin{figure*}
\begin{center}
\includegraphics[width=0.49\textwidth]{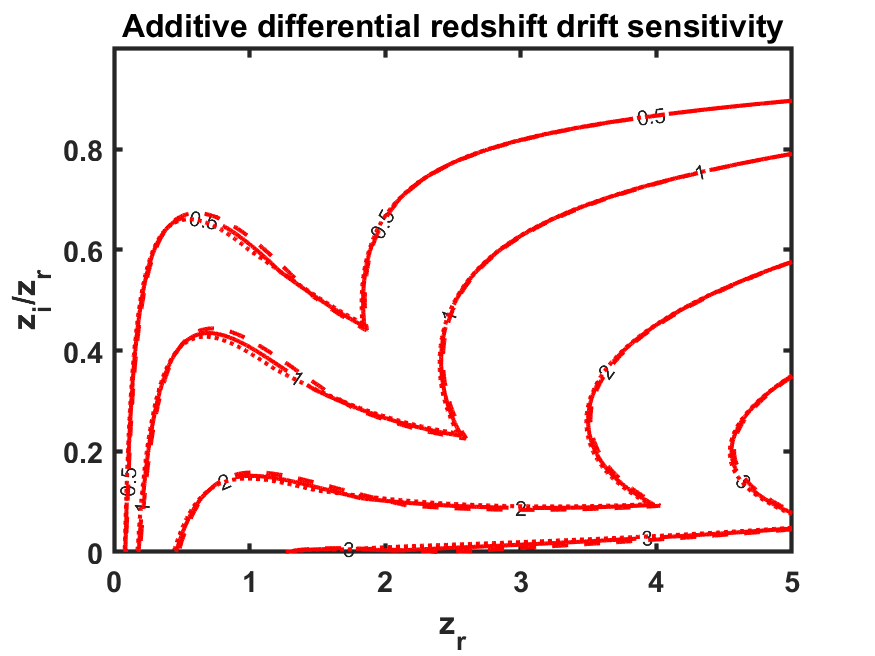}
\includegraphics[width=0.49\textwidth]{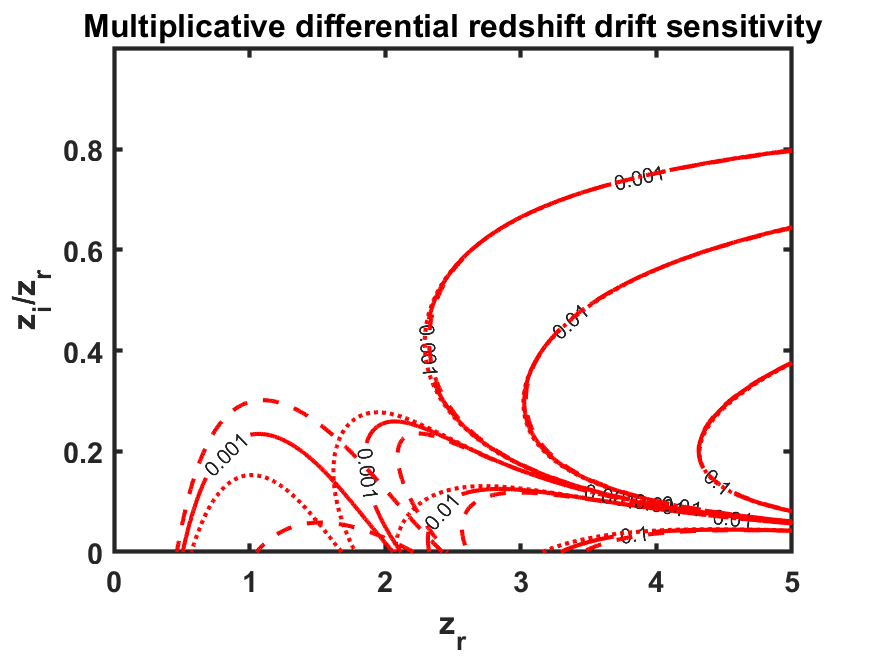}
\includegraphics[width=0.49\textwidth]{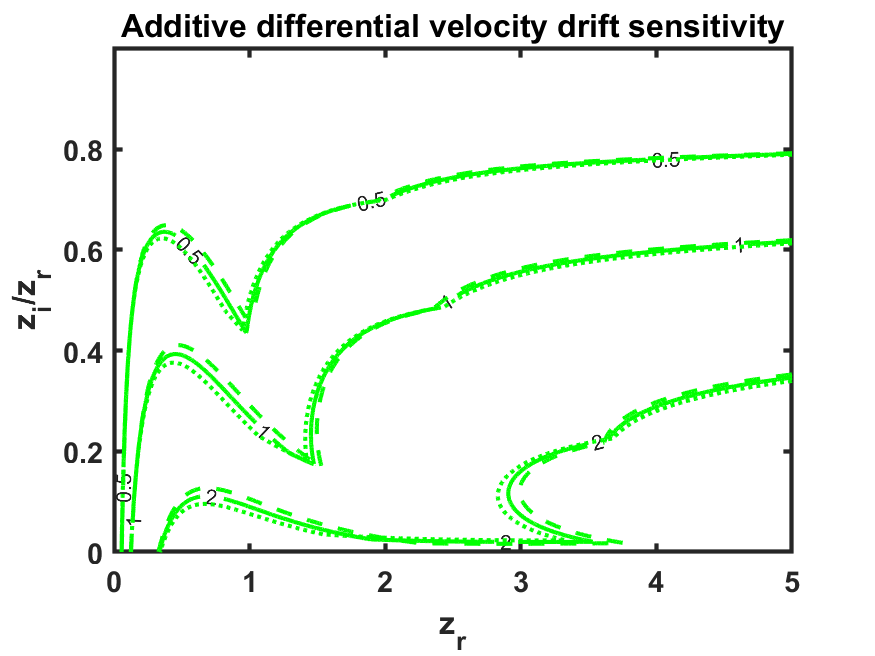}
\includegraphics[width=0.49\textwidth]{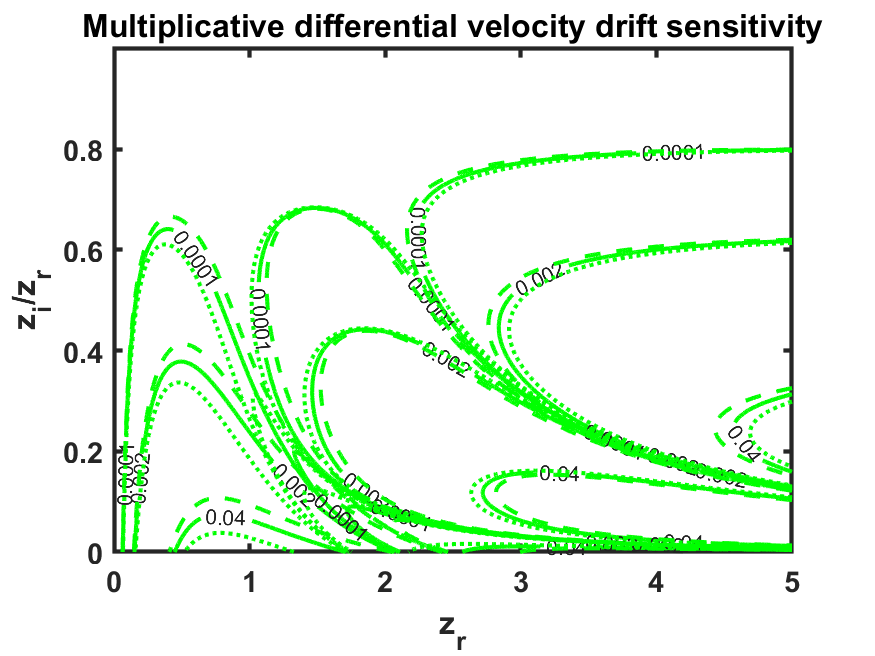}
\end{center}
\caption{Overall additive and multiplicative sensitivities (left and right panels, respectively), for the differential redshift drift (top) and the corresponding spectroscopic velocity (bottom). The solid, dashed, and dotted lines correspond to $\Omega_k=0$, $\Omega_k=-0.1$, and $\Omega_k=+0.1$ respectively.}
\label{figure08}
\end{figure*}

By analogy with the standard redshift drift case, we can also estimate overall sensitivities. As in the former case, we take the absolute values of each of the four sensitivities, normalized to their maximum values in the redshift range under consideration ($0\le z\le5$), and consider, in each case, either the sum or the product of these four sensitivities. These are shown in Figure \ref{figure08}, which is an extension of Figure \ref{figure02}. As in the previous section, we depict selected contour lines of the relevant quantities in the $(z_r,z_i/z_r)$ plane.

The most salient point is that the additive case is rather insensitive to the value of the curvature, while in the multiplicative case there is a considerable sensitivity at low redshifts. For the differential redshift drift itself, the conceptually optimal choice is a reference redshift as large as possible (in our analysis this has been assumed to be $z=5$) and an intervening redshift $z\approx 1$. The same is true for the spectroscopic velocity, although in this case a standard measurement at $z\approx 1$ is a comparable option. Both of these are consistent with the results in the previous sections, highlighting the dual goals of constraining the matter density and the dark energy equation of state.

\section{Global figure-of-merit and joint constraints}
\label{dfom}

We now come to the main point (and the main novelty) of our analysis. For the ELT redshift drift measurements, and specifically for the ANDES Golden Sample, it is quite plausible that the same set of spectra enables a measurement of the standard drift and the differential drift with respect to some non-zero intervening redshift. We provide a first assessment of the constraints which may be obtained from such joint measurements.

For the differential redshift drift there is currently no detailed work proposing a target list analogous to the Golden Sample, though one may surmise that the Golden Sample itself is a good starting point (more on this below). The previous work of \cite{Esteves} explored some of the characteristics which, from a theoretical point of view, such a sample might have. In what follows we provide a more specific forecast, assuming that for each of the seven quasars of the ANDES Golden Sample one can make a measurement of the differential redshift drift in addition to the standard one.

To reduce the parameter space to a manageable size (as well as for the sake of clarity), we assume that all seven measurements of the differential drift are made at the same ratio of intervening and reference redshifts, $z_i/z_r$, and further assume that this ratio can be in the plausible range $z_i/z_r\in[0.15,0.85]$. A further simplifying assumption is that the spectroscopic velocity uncertainty of the differential redshift drift measurement is, for each of the seven targets, proportional to the spectroscopic velocity uncertainty of the standard drift measurement in each quasar; the latter has been estimated in \cite{Cristiani}. Specifically we consider three cases: an optimistic one in which the velocity uncertainty of the differential drift is the same as that of the standard drift, and two more realistic ones where it is worse than it by a factor of two or four. We will assume the same fiducial models and priors as in the previous sections.

\begin{table*}
\begin{center}
\caption{The overall Golden Sample FoM (defined in the main text and rounded off the the nearest integer), for various assumptions on the parameter space and the value of the curvature parameter. The third column has the value for the priors only. The forth has the value for the standard redshift drift plus the priors (together with the FoM gain with respect to the priors only case in the fifth). The sixth contains the FoM for the best-case joint measurement of the two drifts plus the priors (together with the FoM gain with respect to the case without the differential redshift drift in the seventh).}
\label{table5}
\begin{tabular}{c | c | c | c c | c c}
\hline
Parameter space & $\Omega_k$ & Priors only & Standard drift & Gain & Joint (maximal) & Gain \\
\hline
2D $(h,\Omega_m)$ & N/A & 200 & 347 & 1.73 & 427 & 1.23 \\
\hline
3D $(h,\Omega_m,w_0)$ & N/A & 121 & 175 & 1.45 & 208 & 1.19 \\
\hline
4D $(h,\Omega_m,w_0,\Omega_k)$ & -0.1 & 82 & 118 & 1.45 & 138 & 1.17 \\
4D $(h,\Omega_m,w_0,\Omega_k)$ & 0.0 & 82 & 117 & 1.44 & 135 & 1.15 \\
4D $(h,\Omega_m,w_0,\Omega_k)$ & +0.1 & 82 & 116 & 1.42 & 133 & 1.14 \\
\hline
\end{tabular}
\end{center}
\end{table*}

We also need an efficient way to quantify the constraining power of the data, for various assumptions on the model parameter space. The simplest solution is to define, for a generic model with $n$ free parameters, an overall FoM as the inverse nth root of the covariance matrix determinant, FoM$=(det(Cov_n))^{-1/n}$. This serves to quantify both the decreasing constraining power as the parameter space is enlarged and its dependence on the curvature parameter.

Table \ref{table5} summarizes the result of this analysis. We consider the four-dimensional parameter space (with three different fiducial values for $\Omega_k$), but also the reduced three-dimensional case (without the curvature parameter) and even a simpler two-dimensional case, strictly corresponding to flat $\Lambda$CDM. For each of these we show the FoM for the priors only, for their combination with the standard redshift drift measurements, and for the joint measurement of both drifts. In the latter case the table reports the best case, which always corresponds to the lowest choice of the ratio of redshifts,  $z_i/z_r=0.15$, and to equal spectroscopic velocity uncertainties in the two measurements. We find that in the four-dimensional case the standard redshift drift measurements of the Golden Sample improve the priors FoM by a bit more than $40\%$, while an additional measurement of the differential redshift drift at a low intervening redshift improves the said FoM by another $15\%$ or so. We also confirm that the sensitivity is very slightly larger for closed than for open universes.

\begin{figure*}
\begin{center}
\includegraphics[width=0.49\textwidth]{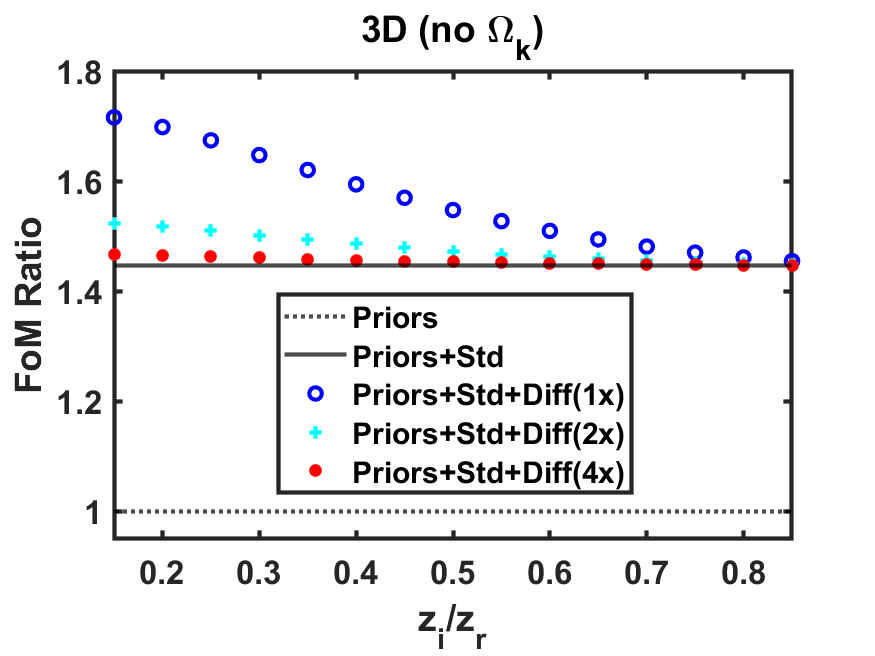}
\includegraphics[width=0.49\textwidth]{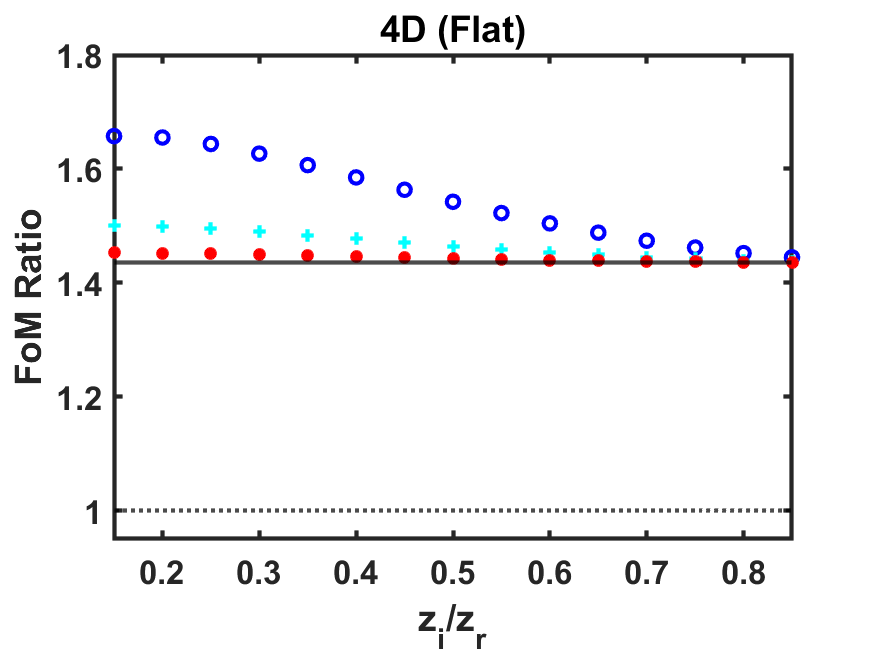}
\includegraphics[width=0.49\textwidth]{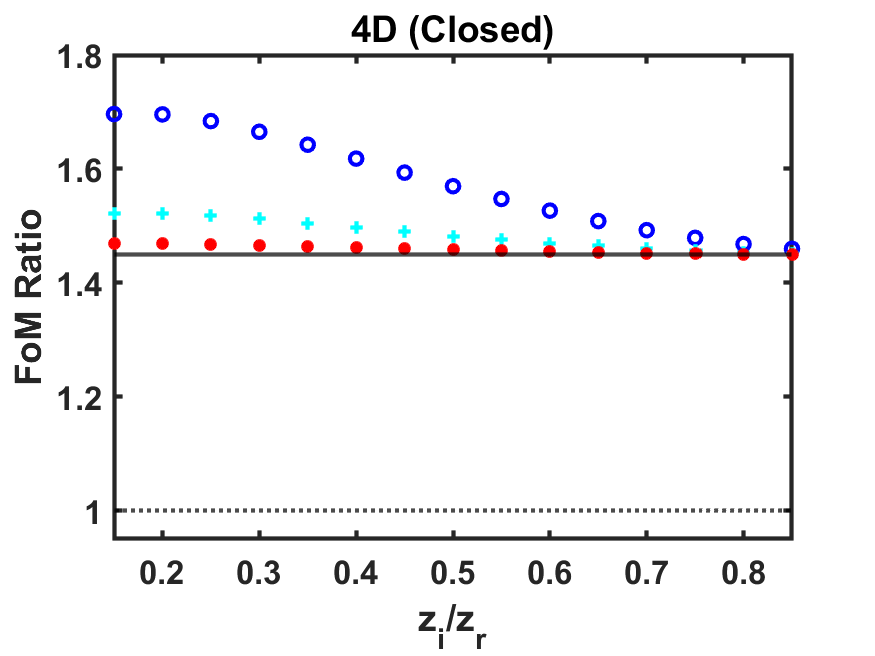}
\includegraphics[width=0.49\textwidth]{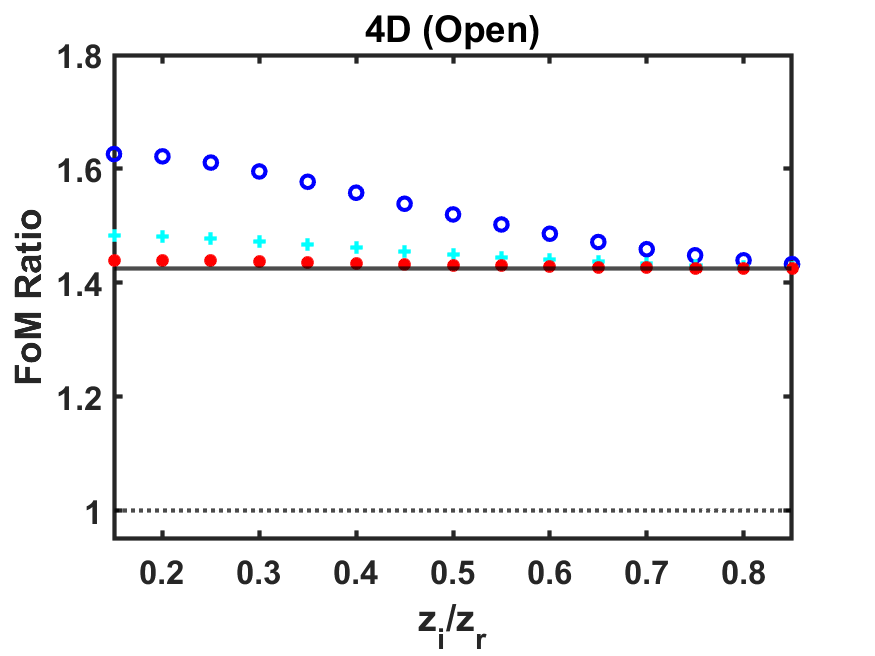}
\end{center}
\caption{The overall Golden Sample FoM (rescaled to the priors' value), for various assumptions on the fiducial model and the measurements. The top left panel is for the three-dimensional parameter space, and the others for four-dimensional parameter space, assuming flat (top right), closed (bottom left), and open (bottom right) universes. The black dashed and solid lines correspond to the priors only, and priors plus standard redshift drift measurements. The color symbols include an additional measurement of the differential redshift drift along each line of sight, at the shown value of $z_i/z_r$, and with a spectroscopic velocity uncertainty that equals that of the standard drift (blue circles) or is larger by a factor of two or four (cyan pluses and red asterisks respectively).}
\label{figure09}
\end{figure*}

Figure \ref{figure09} depicts, for the three and four-dimensional parameter spaces, how the gains afforded by the differential redshift drift measurement depend on the choices of $z_i/z_r$ and its spectroscopic velocity uncertainty, with the previously mentioned assumptions. To facilitate the comparisons, we rescale all FoMs in each case to the value for the priors only. It is interesting that the qualitative behaviour is the same in all cases.

This figure also illustrates two significant points. The first is that the impact of the differential redshift drift measurement is larger for values of $z_i/z_r\approx 0.15$--$0.25$. This is unsurprising given our earlier discussion of the sensitivity to the various model parameters. The second is that the gain in sensitivity brought forth by the differential drift measurement is only significant if its spectroscopic velocity uncertainty is comparable to that of the standard measurement: as the figure shows,  a factor of two difference in this uncertainty removes most of the gains.

\section{Coda: Closing the redshift drift loop}
\label{coda}

Before concluding, we briefly highlight a further advantage of joint measurements of the standard and differential redshift drifts along the same line of sight. Conceptually, the idea is straightforward: if one can directly measure both $S_z(z_r,0)$ and  $S_z(z_r,z_i)$ one can effectively derive a third  measurement since, according to Eq. \ref{difdeltav}
\be
S_v(z_i,0)=S_v(z_r,0)-S_v(z_r,z_i)\,;
\ee
obviously this is an indirect measurement, which is not independent from the other two. However, it is possible that this redshift $z_i$ is sufficiently low to be within the reach of a direct measurement by the SKA. If so, redshift drift measurements using the ANDES Golden Sample could be used to {\it predict, in a model-independent way}, the values of the redshift drift at lower redshift, and such predictions could then be directly tested by the SKA. A confirming result would be a powerful consistency test of our broad cosmological paradigm. As for a disagreement between the two measurements, and setting aside the existence of systematics in one or both measurements, the most natural explanation---at least under the assumption that the lines of sight (or sky patches) would be different for the two measurements---would be a breakdown of the Cosmological Principle, i.e. of the assumption of a homogeneous and isotropic universe \cite{Aluri}.

\begin{figure*}
\begin{center}
\includegraphics[width=0.49\textwidth]{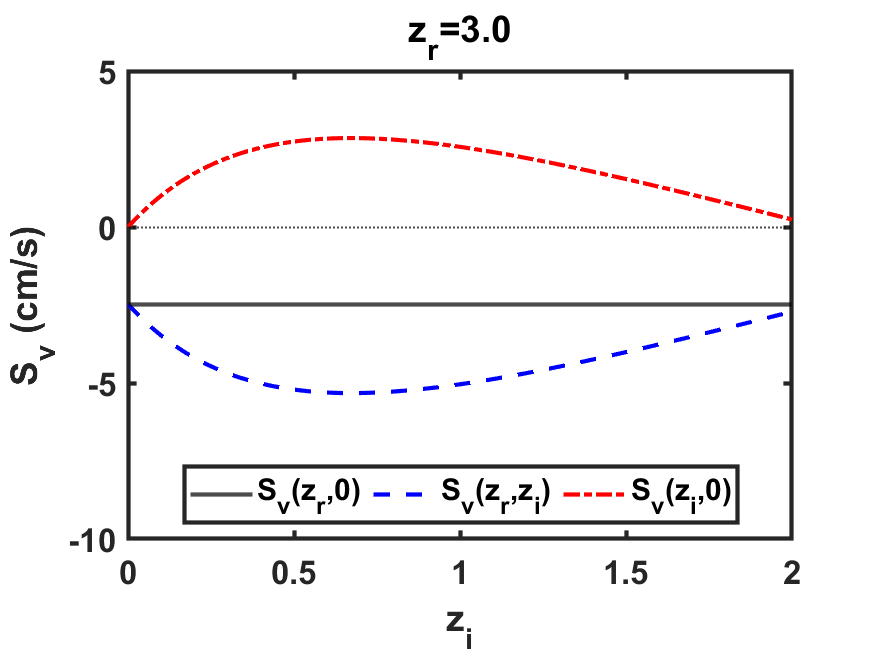}
\includegraphics[width=0.49\textwidth]{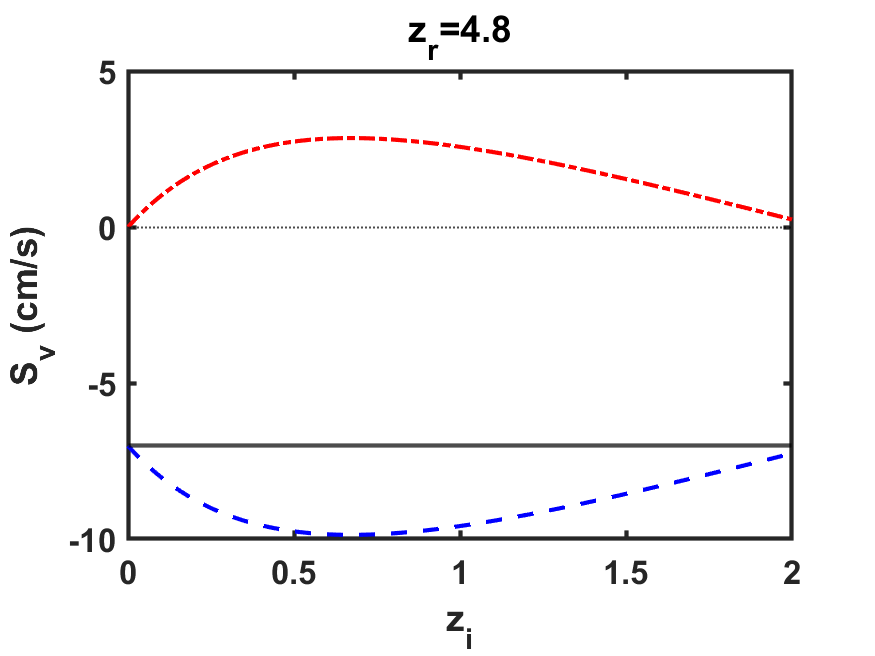}
\end{center}
\caption{The values of the spectroscopic velocities $S_v(z_r,0)$ (black solid), $S_v(z_r,z_i)$ (blue dashed) and $S_v(z_i,0)$ (red dash-dotted) as a function of the intervening redshift $z_i$, for a 10-year experiment time, and the choices of $z_r=3.0$ (left panel) and $z_r=4.8$ (right panel), corresponding to the lowest and highest redshift quasars in the ANDES Golden Sample. For reference, the locus of zero spectroscopic velocity is shown by the thin dotted black line.}
\label{figure10}
\end{figure*}

How feasible such a test is in practice will depend both on the availability of absorption systems at suitable redshifts along the line of sight of the Golden Sample quasars and on the range of redshifts at which the SKA can provide reasonably precise redshift drift measurements. A detailed exploration of these points is left for future work, but here we provide a brief and qualitative illustration of the concept.

The quasars in the Golden Sample, span the approximate redshift range $z_r\in[3.0,4.8]$. At the two ends of this range, and for the same fiducial flat $\Lambda$CDM as assumed elsewhere in this work and an experiment time of 10 years, the predicted spectroscopic velocities are approximately $S_v(3.0,0)=-2.5$ cm/s and $S_v(4.8,0)=-7.0$ cm/s. Figure \ref{figure10} shows, for the two quasars, the expected values $S_v(z_r,z_i)$ and $S_v(z_i,0)$ as a function of the intervening redshift $z_i$. This shows that the ideal range of intervening redshifts for this test is $z_i\in[0.1,1.0]$, which is encouraging both because absorption systems do exist in this redshift range and because direct measurements of the redshift drift in this redshift range should be within the reach of the full SKA. 

\section{Conclusions}
\label{concl}

We have provided an updated assessment of the sensitivity and cosmological impact of forthcoming redshift drift measurements by the ELT and the SKA, including the case of the differential redshift drift (on its own and in combination with the standard measurement) and relaxing the assumption of flat universes. Specific emphasis has been given to the Golden Sample for the ANDES spectrograph \cite{Cristiani}.

Overall, we find that the sensitivity of the redshift drift to curvature is comparable to that of matter (especially at low redshifts) and larger than the sensitivity to the dark energy equation of state, This sensitivity is asymmetric with respect to the curvature parameter, being slightly larger for closed universes than for open universes with the same absolute value of $\Omega_k$.

The simplest way to understand this last result is that in a closed universe there is more dark energy than in an open one---recall that in our fiducial models we are assuming a fixed matter content. As a result the universe starts accelerating earlier (at higher redshifts) in the former case, and consequently the redshift range over which the redshift drift signal (and the corresponding spectroscopic velocity) is positive is larger. The ultimate implication is that at lower redshifts, at which the observables are sensitive to all cosmological parameters, the signal--- whether expressed as $S_v$ or $S_z$---is larger in the closed case. At higher redshifts the signal is smaller, in absolute value, for closed than for open universe (under the same assumptions as above), but at those redshifts the sensitivity to curvature is more indirect: the main direct sensitivity is to the matter density, though this does help with curvature since it can break degeneracies between the two parameters. Admittedly these results can depend on the choice of fiducial model and its parameters. For future work, it will be interesting to check the extent to which these results hold for dynamical dark energy or modified gravity fiducial models.

Focusing on the  Golden Sample, which offers the most promising route for a first detection (and is currently being used for the ESPRESSO redshift drift experiment), we provided a first simple assessment of the impact of joint measurements of the standard and differential redshift drift along the same lines of sight---and therefore at no extra cost in terms of telescope time. Ongoing work in the ESPRESSO experiment, to be reported elsewhere, demonstrates the feasibility of finding absorption features in these spectra which make a differential measurement possible. We also highlighted how such joint measurements enable a third non-independent one, which is conceptually important because it bridges the gap in the redshift ranges directly probed by the ELT and the SKA.

It remains to be seen how the uncertainty achievable for differential redshift drift measurements compares to that of the standard drift measurements in the same spectra, and at what redshifts these can be done. As our analysis confirms, having comparable sensitivities and intervening redshifts in the approximate redshift range $z_i\approx 0.75$--$1.25$ would be highly desirable. In any case, provided the ANDES precision and stability  \cite{ANDES} are sufficient to fully exploit the ELT photons (rather than being limited by systematics), the measurement of both redshift drifts is a realistic possibility.

\section*{Data availability}

This work uses simulated data, generated using the procedures detailed in the text.

\section*{Acknowledgements}

We acknowledge useful discussions on the topic of this work with Stefano Cristiani, Joe Liske and Andrea Trost.

This work was financed by Portuguese funds through FCT (Funda\c c\~ao para a Ci\^encia e a Tecnologia) in the framework of the project 2022.04048.PTDC (Phi in the Sky, DOI 10.54499/2022.04048.PTDC). CJM also acknowledges FCT and POCH/FSE (EC) support through Investigador FCT Contract 2021.01214.CEECIND/CP1658/CT0001 (DOI 10.54499/2021.01214.CEECIND/CP1658/CT0001). SQF was partially supported by Ci\^encia
Viva OCJF funds. CMJM is supported by an FCT fellowship, grant number 2023.03984.BD.

\bibliographystyle{model1-num-names}
\bibliography{drift}

\begin{thebibliography}{21}
\expandafter\ifx\csname natexlab\endcsname\relax\def\natexlab#1{#1}\fi
\providecommand{\bibinfo}[2]{#2}
\ifx\xfnm\relax \def\xfnm[#1]{\unskip,\space#1}\fi
\bibitem[{{Sandage}(1962)}]{Sandage}
\bibinfo{author}{A.~{Sandage}},
\newblock \bibinfo{title}{{The Change of Redshift and Apparent Luminosity of Galaxies due to the Deceleration of Selected Expanding Universes.}},
\newblock \bibinfo{journal}{Astrophys. J.} \bibinfo{volume}{136} (\bibinfo{year}{1962}) \bibinfo{pages}{319}.
\bibitem[{Liske et~al.(2008)}]{Liske}
\bibinfo{author}{J.~Liske}, et~al.,
\newblock \bibinfo{title}{{Cosmic dynamics in the era of Extremely Large Telescopes}},
\newblock \bibinfo{journal}{Mon. Not. Roy. Astron. Soc.} \bibinfo{volume}{386} (\bibinfo{year}{2008}) \bibinfo{pages}{1192--1218}.
\bibitem[{Uzan et~al.(2008)Uzan, Clarkson, and Ellis}]{Uzan}
\bibinfo{author}{J.-P. Uzan}, \bibinfo{author}{C.~Clarkson}, \bibinfo{author}{G.~F.~R. Ellis},
\newblock \bibinfo{title}{{Time drift of cosmological redshifts as a test of the Copernican principle}},
\newblock \bibinfo{journal}{Phys. Rev. Lett.} \bibinfo{volume}{100} (\bibinfo{year}{2008}) \bibinfo{pages}{191303}.
\bibitem[{Quercellini et~al.(2012)Quercellini, Amendola, Balbi, Cabella, and Quartin}]{Quercellini}
\bibinfo{author}{C.~Quercellini}, \bibinfo{author}{L.~Amendola}, \bibinfo{author}{A.~Balbi}, \bibinfo{author}{P.~Cabella}, \bibinfo{author}{M.~Quartin},
\newblock \bibinfo{title}{{Real-time Cosmology}},
\newblock \bibinfo{journal}{Phys. Rept.} \bibinfo{volume}{521} (\bibinfo{year}{2012}) \bibinfo{pages}{95--134}.
\bibitem[{Heinesen(2021)}]{Heinesen}
\bibinfo{author}{A.~Heinesen},
\newblock \bibinfo{title}{{Redshift drift as a model independent probe of dark energy}},
\newblock \bibinfo{journal}{Phys. Rev. D} \bibinfo{volume}{103} (\bibinfo{year}{2021}) \bibinfo{pages}{L081302}.
\bibitem[{Darling(2012)}]{Darling}
\bibinfo{author}{J.~Darling},
\newblock \bibinfo{title}{{Toward a Direct Measurement of the Cosmic Acceleration}},
\newblock \bibinfo{journal}{Astrophys. J.} \bibinfo{volume}{761} (\bibinfo{year}{2012}) \bibinfo{pages}{L26}.
\bibitem[{Cooke(2020)}]{Cooke}
\bibinfo{author}{R.~Cooke},
\newblock \bibinfo{title}{{The ACCELERATION programme: I. Cosmology with the redshift drift}},
\newblock \bibinfo{journal}{Mon. Not. Roy. Astron. Soc.} \bibinfo{volume}{492} (\bibinfo{year}{2020}) \bibinfo{pages}{2044--2057}.
\bibitem[{Kl\"ockner et~al.(2015)Kl\"ockner, Obreschkow, Martins, Raccanelli, Champion, Roy, Lobanov, Wagner, and Keller}]{Klockner}
\bibinfo{author}{H.-R. Kl\"ockner}, \bibinfo{author}{D.~Obreschkow}, \bibinfo{author}{C.~Martins}, \bibinfo{author}{A.~Raccanelli}, \bibinfo{author}{D.~Champion}, \bibinfo{author}{A.~L. Roy}, \bibinfo{author}{A.~Lobanov}, \bibinfo{author}{J.~Wagner}, \bibinfo{author}{R.~Keller},
\newblock \bibinfo{title}{{Real time cosmology - A direct measure of the expansion rate of the Universe with the SKA}},
\newblock \bibinfo{journal}{PoS} \bibinfo{volume}{AASKA14} (\bibinfo{year}{2015}) \bibinfo{pages}{027}.
\bibitem[{Liske et~al.(2014)}]{HIRES}
\bibinfo{author}{J.~Liske}, et~al., \bibinfo{title}{{Top Level Requirements For ELT-HIRES}}, \bibinfo{type}{Technical Report}, Document ESO 204697 Version 1, \bibinfo{year}{2014}.
\bibitem[{Martins et~al.(2024)}]{ANDES}
\bibinfo{author}{C.~J. A.~P. Martins}, et~al.,
\newblock \bibinfo{title}{{Cosmology and fundamental physics with the ELT-ANDES spectrograph}},
\newblock \bibinfo{journal}{Exper. Astron.} \bibinfo{volume}{57} (\bibinfo{year}{2024}) \bibinfo{pages}{5}.
\bibitem[{Alves et~al.(2019)Alves, Leite, Martins, Matos, and Silva}]{Alves}
\bibinfo{author}{C.~S. Alves}, \bibinfo{author}{A.~C.~O. Leite}, \bibinfo{author}{C.~J. A.~P. Martins}, \bibinfo{author}{J.~G.~B. Matos}, \bibinfo{author}{T.~A. Silva},
\newblock \bibinfo{title}{{Forecasts of redshift drift constraints on cosmological parameters}},
\newblock \bibinfo{journal}{Mon. Not. Roy. Astron. Soc.} \bibinfo{volume}{488} (\bibinfo{year}{2019}) \bibinfo{pages}{3607--3624}.
\bibitem[{Rocha and Martins(2022)}]{Rocha}
\bibinfo{author}{B.~A.~R. Rocha}, \bibinfo{author}{C.~J. A.~P. Martins},
\newblock \bibinfo{title}{{Redshift drift cosmography with ELT and SKAO measurements}},
\newblock \bibinfo{journal}{Mon. Not. Roy. Astron. Soc.} \bibinfo{volume}{518} (\bibinfo{year}{2022}) \bibinfo{pages}{2853--2869}.
\bibitem[{Marques et~al.(2023)Marques, Martins, and L\'opez}]{SKA}
\bibinfo{author}{C.~M.~J. Marques}, \bibinfo{author}{C.~J. A.~P. Martins}, \bibinfo{author}{B.~G. L\'opez},
\newblock \bibinfo{title}{{Watching the Universe\textquoteright{}s acceleration era with the SKAO}},
\newblock \bibinfo{journal}{Mon. Not. Roy. Astron. Soc.} \bibinfo{volume}{527} (\bibinfo{year}{2023}) \bibinfo{pages}{9918--9929}.
\bibitem[{Balbi and Quercellini(2007)}]{Balbi}
\bibinfo{author}{A.~Balbi}, \bibinfo{author}{C.~Quercellini},
\newblock \bibinfo{title}{{The time evolution of cosmological redshift as a test of dark energy}},
\newblock \bibinfo{journal}{Mon. Not. Roy. Astron. Soc.} \bibinfo{volume}{382} (\bibinfo{year}{2007}) \bibinfo{pages}{1623--1629}.
\bibitem[{Aghanim et~al.(2020)}]{Planck}
\bibinfo{author}{N.~Aghanim}, et~al.,
\newblock \bibinfo{title}{{Planck 2018 results. VI. Cosmological parameters}},
\newblock \bibinfo{journal}{Astron. Astrophys.} \bibinfo{volume}{641} (\bibinfo{year}{2020}) \bibinfo{pages}{A6}. \bibinfo{note}{[Erratum: Astron.Astrophys. 652, C4 (2021)]}.
\bibitem[{Di~Valentino et~al.(2019)Di~Valentino, Melchiorri, and Silk}]{Closed}
\bibinfo{author}{E.~Di~Valentino}, \bibinfo{author}{A.~Melchiorri}, \bibinfo{author}{J.~Silk},
\newblock \bibinfo{title}{{Planck evidence for a closed Universe and a possible crisis for cosmology}},
\newblock \bibinfo{journal}{Nature Astron.} \bibinfo{volume}{4} (\bibinfo{year}{2019}) \bibinfo{pages}{196--203}.
\bibitem[{Cristiani et~al.(2023)}]{Cristiani}
\bibinfo{author}{S.~Cristiani}, et~al.,
\newblock \bibinfo{title}{{Spectroscopy of QUBRICS quasar candidates: 1672 new redshifts and a golden sample for the Sandage test of the redshift drift}},
\newblock \bibinfo{journal}{Mon. Not. Roy. Astron. Soc.} \bibinfo{volume}{522} (\bibinfo{year}{2023}) \bibinfo{pages}{2019--2028}.
\bibitem[{Dong et~al.(2022)Dong, Gonzalez, Eikenberry, Jeram, Likamonsavad, Liske, Stelter, and Townsend}]{Dong}
\bibinfo{author}{C.~Dong}, \bibinfo{author}{A.~Gonzalez}, \bibinfo{author}{S.~Eikenberry}, \bibinfo{author}{S.~Jeram}, \bibinfo{author}{M.~Likamonsavad}, \bibinfo{author}{J.~Liske}, \bibinfo{author}{D.~Stelter}, \bibinfo{author}{A.~Townsend},
\newblock \bibinfo{title}{{Forecasting cosmic acceleration measurements using the Lyman-\ensuremath{\alpha} forest}},
\newblock \bibinfo{journal}{Mon. Not. Roy. Astron. Soc.} \bibinfo{volume}{514} (\bibinfo{year}{2022}) \bibinfo{pages}{5493--5505}.
\bibitem[{{Marques} et~al.(2023){Marques}, {Martins}, and {Alves}}]{FRIDDA}
\bibinfo{author}{C.~M.~J. {Marques}}, \bibinfo{author}{C.~J.~A.~P. {Martins}}, \bibinfo{author}{C.~S. {Alves}}, \bibinfo{title}{{FRIDDA: Fisher foRecast code for combIned reDshift Drift and Alpha}}, \bibinfo{howpublished}{Astrophysics Source Code Library, record ascl:2305.001}, \bibinfo{year}{2023}.
\bibitem[{Esteves et~al.(2021)Esteves, Martins, Pereira, and Alves}]{Esteves}
\bibinfo{author}{J.~Esteves}, \bibinfo{author}{C.~J. A.~P. Martins}, \bibinfo{author}{B.~G. Pereira}, \bibinfo{author}{C.~S. Alves},
\newblock \bibinfo{title}{{Cosmological impact of redshift drift measurements}},
\newblock \bibinfo{journal}{Mon. Not. Roy. Astron. Soc.} \bibinfo{volume}{508} (\bibinfo{year}{2021}) \bibinfo{pages}{L53--L57}.
\bibitem[{Aluri et~al.(2023)}]{Aluri}
\bibinfo{author}{P.~K. Aluri}, et~al.,
\newblock \bibinfo{title}{{Is the observable Universe consistent with the cosmological principle?}},
\newblock \bibinfo{journal}{Class. Quant. Grav.} \bibinfo{volume}{40} (\bibinfo{year}{2023}) \bibinfo{pages}{094001}.

\end{thebibliography}
\end{document}